\documentclass[aps,superscriptaddress,twocolumn]{revtex4-1}

\pdfoutput=1

\usepackage{graphicx}
\usepackage{amsmath}
\usepackage{amssymb}
\usepackage{verbatim}
\usepackage{braket}
\usepackage{color}
\usepackage{hyperref}
\hypersetup{
	colorlinks=true,
	linkcolor=red,      
	urlcolor=blue,
	citecolor=blue
}

\begin{document}
\title{Signatures of topological phase transition on a quantum critical line}
	
\author{Ranjith R Kumar}
\affiliation{Theoretical Sciences Division, Poornaprajna Institute of Scientific Research, Bidalur, Bengaluru-562164, India.}
\affiliation{Graduate Studies, Manipal Academy of Higher Education, Madhava Nagar, Manipal-576104, India.}
\author{Nilanjan Roy}
\affiliation{Centre for Condensed Matter Theory, Department of Physics, Indian Institute of Science, Bengaluru-560012, India.}
\author{Y R Kartik}
\affiliation{Theoretical Sciences Division, Poornaprajna Institute of Scientific Research, Bidalur, Bengaluru-562164, India.}
\affiliation{Graduate Studies, Manipal Academy of Higher Education, Madhava Nagar, Manipal-576104, India.}
\author{S Rahul}
\affiliation{Theoretical Sciences Division, Poornaprajna Institute of Scientific Research, Bidalur, Bengaluru-562164, India.}
\affiliation{Graduate Studies, Manipal Academy of Higher Education, Madhava Nagar, Manipal-576104, India.}
\author{Sujit Sarkar}
\affiliation{Theoretical Sciences Division, Poornaprajna Institute of Scientific Research, Bidalur, Bengaluru-562164, India.}
\date{\today}

\begin{abstract}
Recently topological states of matter have witnessed a new physical phenomenon where both edge modes and gapless bulk coexist at topological quantum criticality.
The presence and absence of edge modes on a critical line can lead to an unusual class of topological phase transition between the topological and non-topological critical phases. 
We explore the existence of this new class of topological phase transitions in a generic model representing the topological insulators and superconductors and we show that such transition occurs 
at a multicritical point i.e. at the intersection of two critical lines. To characterize these transitions we reconstruct the theoretical frameworks which include bound state solution of the Dirac equation, winding number, correlation factors and
scaling theory of the curvature function to work for the criticality. Critical exponents and scaling laws are discussed to distinguish between the multicritical points which separate the critical phases. Entanglement entropy and its scaling in the real-space provide further insights into the unique transition at criticality revealing the interplay between fixed point and critical point at the multicriticalities.
\end{abstract}

\maketitle

\section{Introduction}
In the quest of classifying novel phases of quantum matter in the absence of local order parameters, the topology of electronic band structure plays a prime role \cite{haldane1988model,kane2005quantum,narang2021topology,wang2017topological}. It
enables the distinction
between gapped phases in terms of a quantized invariant number, which counts the number of localized edge modes present \cite{thouless1982quantized}. The transition between the distinct topological phases involves a bulk band gap closing at the point of topological phase transition. Across the transition the number of edge modes, protected by the bulk gap, changes \cite{hasan2010colloquium}. In the gapped phases, the localization length of the edge modes diverges as the system drives towards the transition point or criticality \cite{continentino2020finite}.

Interestingly, this conventional knowledge is revised recently, realizing even criticalities can host the stable localized edge modes despite the vanishing bulk gap \cite{verresen2018topology,verresen2019gapless,jones2019asymptotic,verresen2020topology,rahul2021majorana,niu2021emergent,PhysRevB.104.075132,PhysRevResearch.3.043048,fraxanet2021topological,keselman2015gapless,scaffidi2017gapless,duque2021topological}. 
This results in the emergence of non-trivial criticalities with unique topological properties even in the presence of gapless bulk excitations.  
The non-trivial criticalities can be effectively characterized in terms of the zeros and poles of complex function associated with the Hamiltonian \cite{verresen2018topology}.
The localized edge 
modes at criticality are 
protected by novel phenomena such as kinetic inversion \cite{verresen2020topology} (in fermionic models) and finite high energy charge gap \cite{fraxanet2021topological} (in bosonic models).
It has been shown that they also remain  robust against interactions and disorders \cite{verresen2018topology,verresen2019gapless}. 
This intriguing interplay between topology and criticality 
causes an unconventional topological transition between critical phases  \cite{verresen2020topology,kumar2021multi}.

In this work, we report the possibility of a new kind of topological phase transition between critical phases, happening at a multicritical point where two critical lines intersect. Considering a two-band model as a prototype representing gapless (critical) topological insulators and superconductors, in Sec.~\ref{sec2}, we find the critical lines in the model and existence of two species of multicritical points with quadratic and linear dispersions, both corresponding to the topological transition between distinct critical phases. In support of our finding, in Sec.~\ref{sec3}, we solve the Dirac equation for criticality to obtain the localized edge mode solutions in the non-trivial critical phases. This is further supported by our proposal of obtaining the non-zero integer winding number 
for non-trivial critical phases which we discuss in Sec.~\ref{sec4a}. This proposal is supported in Sec.~\ref{sec4b} by calculating the winding number using zeros and poles of a complex function. The curvature function diverges on approaching the multicritical points from critical phases indicating the existence of a transtion between critical phases.
The critical exponents $(\gamma,\nu,z)$, discussed in Sec.~\ref{expo}, unravel the different universality classes of two multicritical points. 
Using the divergence of curvature function we develop a renormalization group (RG) scheme to distinguish the different critical phases in Sec.~\ref{sec5b}. In Sec.~\ref{sec5c} we show that a correlation length, extracted from the Fourier-transformed curvature function, diverges at the multicritical points indicating phase transitions between critical phases. Further, in Sec.~\ref{sec6} we study the spatial scaling of entanglemet entropy, especially to characterize and distinguish the multicritical points. The scaling reveals an interplay between the fixed points and multicritical points. The entanglement entropy is minimum where the fixed point and multicritical point overlap reflecting the dominance of the former over the latter. Finally, we conclude in Sec.~\ref{sec7} .

\section{Model}\label{sec2}
We consider a one-dimensional lattice chain of
spinless fermions in momentum space \cite{PhysRevLett.42.1698,kitaev2001unpaired} represented by a generic two-band Bloch Hamiltonian of the form 
\begin{eqnarray}
\mathcal{H}(k,\boldsymbol{\Gamma}) = \boldsymbol{\chi.\sigma} = \chi_{x} \sigma_x + \chi_{y} \sigma_y,
\label{generic-H}
\end{eqnarray}
where $\boldsymbol{\Gamma}=\left\lbrace \Gamma_{0},\Gamma_{1},\Gamma_{2}\right\rbrace $, $ \chi_{x} = \Gamma_0 + \Gamma_1 \cos k + \Gamma_2 \cos 2k,$ and $ \chi_{y} = \Gamma_1 \sin k + \Gamma_2 \sin 2k$ and $\boldsymbol{\sigma}=(\sigma_x,\sigma_y)$ are the Pauli matrices. 
The model represents extended Su-Schrieffer-Heeger (SSH) \cite{hsu2020topological} and extended Kitaev models \cite{niu2012majorana} by uniquely 
defining the parameters (see Appendix.\ref{Model_Supply} for detailed discussion on the physical relevance of the model). The parameters $\Gamma_{0}$, $\Gamma_{1}$ and $\Gamma_{2}$ describe the onsite potential, 
the nearest neighbor (NN) couplings and the next nearest neighbor (NNN) couplings respectively.
\begin{figure}[t]
	\centering 
	\includegraphics[width=\columnwidth,height=3.8cm]{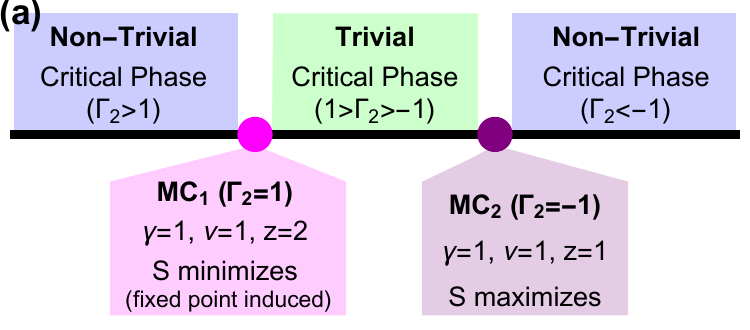}\vspace{0.7cm}
	\includegraphics[width=\columnwidth,height=3.8cm]{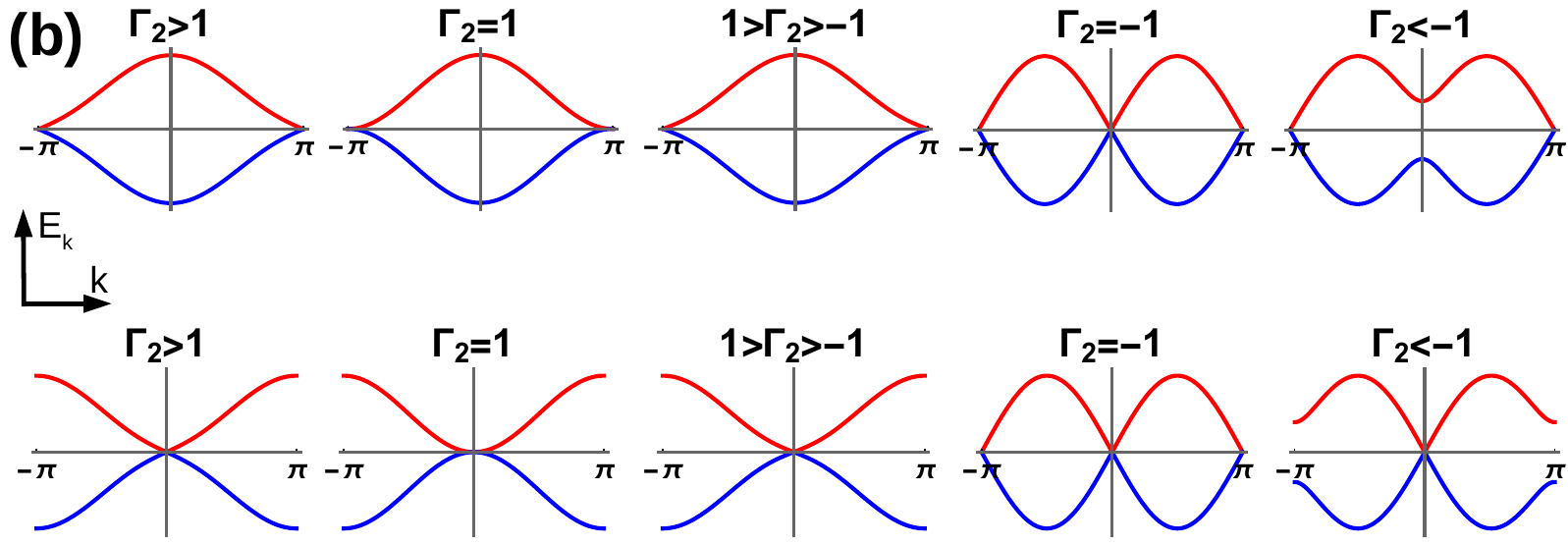}
	\caption{\small \label{Model}(a) Schematic representation of both the criticalities $\Gamma_{1}=\pm(\Gamma_{0}+\Gamma_{2})$ with $\Gamma_{0}=1$. Non-trivial critical phases: $1<\Gamma_{2}$ and $\Gamma_{2}<-1$ (solid lines), trivial critical phase: $1>\Gamma_{2}>-1$ (dashed line) are separated by the multicriticalities $MC_{1,2}$. The entanglement entropy ($S$) minimizes at $MC_1$, which is also an RG fixed point, and maximizes at $MC_2$. The $MC_{1,2}$ belongs to the universality classes obtained from the exponents ($\gamma$,$\nu$,$z$)=(1,1,2) and ($\gamma$,$\nu$,$z$)=(1,1,1) respectively. (b) Dispersion at different critical phases and multicritical points. Upper panel: on the critical line for $k_0=\pm \pi$. Lower panel: on the critical line for $k_0=0$.}
\end{figure} 

In general, the model  
can support three distinct gapped phases distinguished by the number of edge modes they possess. These phases can be identified with the winding numbers $w=0,1$ and $2$, which quantify the edge modes. The model undergoes transition between these phases with necessarily involving the gap closing, $E_k=\pm \sqrt{\chi_{x}^2+\chi_{y}^2}=0$, at the phase boundaries. 
The criticalities, where the bulk gap closes, occurs for the momentum $k_0=0,\pm\pi,\cos^{-1}(-\Gamma_1/2\Gamma_2)$, which respectively gives the critical surfaces $\Gamma_{1}=-(\Gamma_{0}+\Gamma_{2})$, $\Gamma_{1}=(\Gamma_{0}+\Gamma_{2})$ and $\Gamma_{0}=\Gamma_{2}$. The model possesses three multicritical lines at which two critical surfaces intersect. Two of them ($MC_1$) share identical properties and shows quadratic dispersion around the gap closing point and the other one ($MC_2$) is identified with the linear dispersion. Uniquely, the model supports the edge modes and topological transition at critical surfaces $\Gamma_1=\pm(\Gamma_0+ \Gamma_2)$ corresponding to $k_0=\pi,0$ respectively (see Appendix.\ref{Numerical} for the numerical results). 

To further explore these unique phenomena we propose a framework that works out for criticality without referring to any 
of the gapped phases of the model. 
Ideally driving the system to criticality involves $k\rightarrow k_0$ and $\boldsymbol{\Gamma} \rightarrow \boldsymbol{\Gamma}_c$, where $\boldsymbol{\Gamma}_c$ is the critical point in the parameter space. To avoid the singularities involving the exact critical point, one can define
the Hamiltonian critical only in the parameter space as $\mathcal{H}(k,\boldsymbol{\Gamma}_c)$ with $k = k_0 + \Delta k$, where $\Delta k<<2\pi$. This situation is hereafter referred as \textit{criticality} in this work. 

The model at criticality can be obtained by using the critical surface relation which modifies the components into $ \chi_{x} =\Gamma_0 (1+ \cos k) + \Gamma_2 (\cos 2k + \cos k),$ and $ \chi_{y} = \Gamma_2 (\sin 2k + \sin k)+ \Gamma_0 \sin k$ for $\Gamma_{1}=(\Gamma_{0}+\Gamma_{2})$ and $ \chi_{x} =\Gamma_0 (1- \cos k) + \Gamma_2 (\cos 2k - \cos k),$ and $ \chi_{y} = \Gamma_2 (\sin 2k - \sin k)- \Gamma_0 \sin k$ for $\Gamma_{1}=-(\Gamma_{0}+\Gamma_{2})$. The possible topological trivial and non-trivial critical phases are separated by the phase boundaries 
at the multicritical lines $\Gamma_2=\Gamma_0$ ($MC_1$) and $\Gamma_2=-\Gamma_0$ ($MC_2$). Without loss of any generality, we assume $\Gamma_0=1$. Hence hereafter the critical surfaces and the multicritical lines will be called as the critical lines and the multicritical points respectively, as shown in Fig.~\ref{Model}(a). The multicriticalities, $MC_{1,2}$, are identified with quadratic and linear dispersion respectively, as shown in Fig.~\ref{Model}(b). They can be obtained for the following $k_0^{mc}$.
For $\Gamma_2=\Gamma_0$ ($MC_1$):
\begin{align}
k_0^{mc}&=\cos^{-1}\left( -\frac{\Gamma_2+\Gamma_0}{2\Gamma_2}\right)  \quad \text{at} \quad \Gamma_{1}=(\Gamma_{0}+\Gamma_{2}),\\
k_0^{mc}&=\cos^{-1}\left( \frac{\Gamma_2+\Gamma_0}{2\Gamma_2}\right)  \quad \text{at} \quad \Gamma_{1}=-(\Gamma_{0}+\Gamma_{2}).
\end{align} 
For $\Gamma_2=-\Gamma_0$ ($MC_2$):
\begin{align}
k_0^{mc}&=0 \quad \text{at} \quad \Gamma_{1}=(\Gamma_{0}+\Gamma_{2}), \label{swap-1}\\
k_0^{mc}&=\pi \quad \text{at} \quad \Gamma_{1}=-(\Gamma_{0}+\Gamma_{2}). \label{swap-2}
\end{align} 
Interestingly, $MC_2$ exhibits \textit{swapping} of the values of $k_0^{mc}$.
At $MC_2$, one can observe that $k_0^{mc}=0$ for $\Gamma_{1}=(\Gamma_{0}+\Gamma_{2})$ which was obtained for $k_0=\pi$ and $k_0^{mc}=\pi$ for $\Gamma_{1}=-(\Gamma_{0}+\Gamma_{2})$ which was obtained for $k_0=0$. This property emerge as a result of intersection of both the critical lines at $MC_2$.
We will show in the following that our proposed framework based on the near-critical Hamiltonian  $\mathcal{H}(k,\boldsymbol{\Gamma}_c)$ can capture the essential physics of topological transition at criticality.

\section{Bound state solution of the Dirac equation}\label{sec3}
\begin{figure}[t]
		\includegraphics[width=0.45\columnwidth,height=2.5cm]{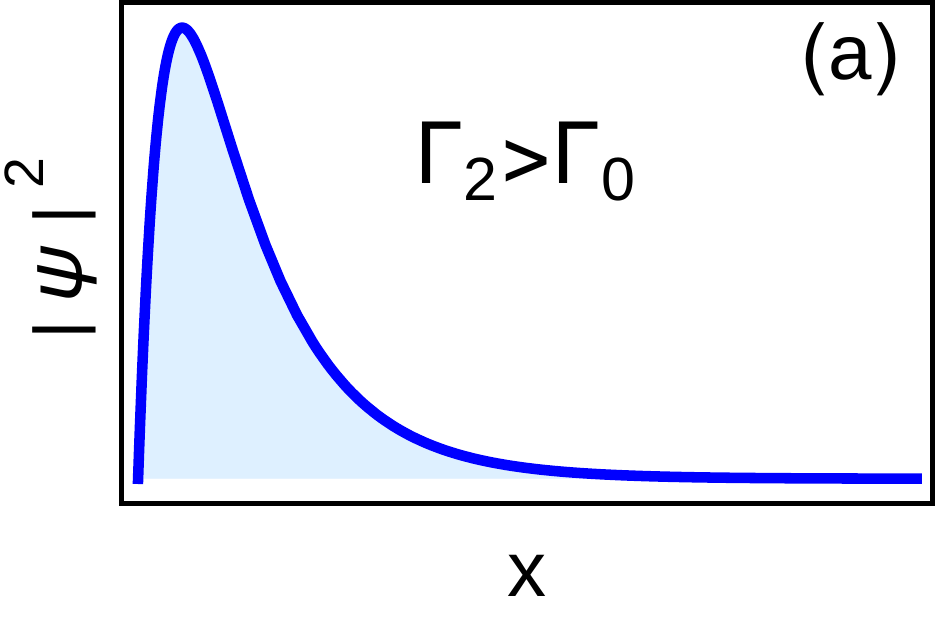}  
	\hspace{0.15cm}
	\includegraphics[width=0.45\columnwidth,height=2.5cm]{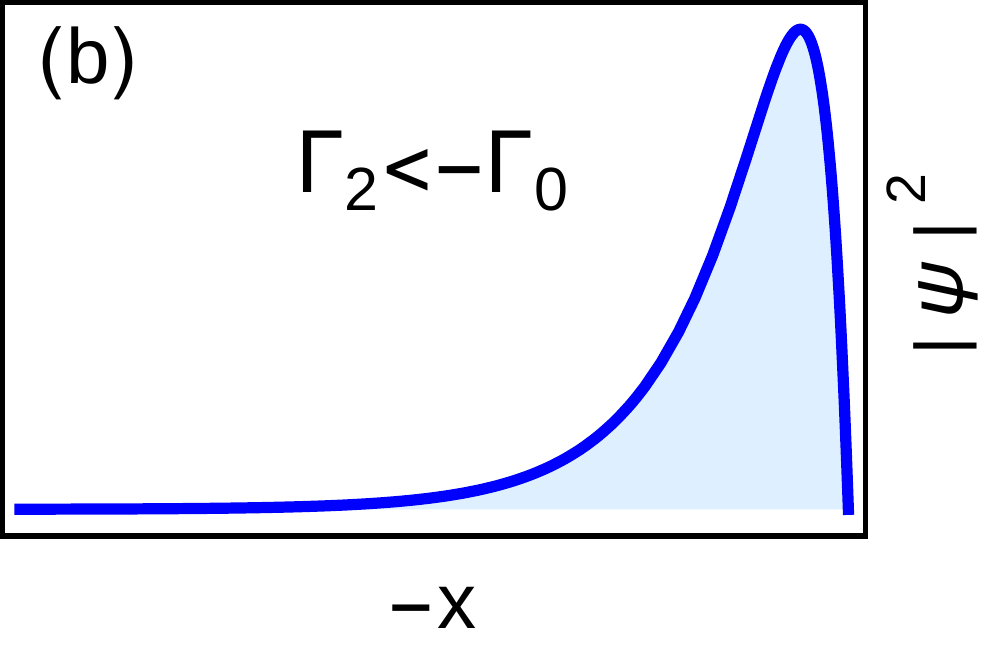} \hspace{0.15cm} 
	\caption{\small \label{Dirac}
		Bound state solutions of the edge modes at the non-trivial critical phases. (a) Plotted for $\Gamma_{2}=3 \Gamma_{0}$ (with $\Gamma_{0}=1$) at the critical phase $\Gamma_{2}>\Gamma_{0}$. (b) Plotted for $\Gamma_{2}=-3 \Gamma_{0}$ (with $\Gamma_{0}=1$) at the critical phase $\Gamma_{2}<-\Gamma_{0}$.}
\end{figure}

The presence of edge modes in topological insulators and superconductors is lucid
from the bound state solution of Dirac equation \cite{jackiw1976solitons,lu2011non,shun2018topological}. 
We solve the model in Eq.\ref{generic-H} for the bound state solution at criticality (see Appendix.\ref{dirac-gapped} for the bound state solution at gapped phases).
Interestingly, as a consequence of the near-critical approach adopted here, 
the Dirac Hamiltonian at criticality naturally fixes the  interface at a multicritical point.
Dirac Hamiltonian at criticality up to third-order expansion around $k_0^{mc}$ for $MC_1$ is
\begin{equation}
\mathcal{H}(k) \approx \epsilon_1 k^2\sigma_x +  (\epsilon_2 k - \epsilon_3 k^3) \sigma_y.
\end{equation}
where $\epsilon_1=(\Gamma_{0}-3\Gamma_{2})/2$, $\epsilon_2=(\Gamma_{2}-\Gamma_{0})$ and $\epsilon_3= (7\Gamma_{2}-\Gamma_{0})/6$. We look for zero energy solution in real space (we set $\hbar=1$ throughout the discussion), $\mathcal{H}
\psi(x)=0$. Identifying the spinor of the wave-function $\psi(x)=\rho_{\eta} \phi(x)$ is an eigenstate of $\sigma_z$ 
and 
using $\phi(x)\propto e^{-x/\xi}$, inverse of the non-zero decay length can be obtained as $\xi^{-1}_{\pm}=(-\eta \epsilon_1 \pm \sqrt{\epsilon_1^2-4\epsilon_2\epsilon_3})/(-2\epsilon_3)$.
For both roots to be positive, it requires $\xi^{-1}_+ + \xi^{-1}_->0$, which implies $\eta=\text{sign}(\epsilon_1/\epsilon_3)$. 
The edge mode 
decay length (longer one of two) is $\xi_+\approx |\epsilon_1|/ \epsilon_2 $ remains finite and positive for $\epsilon_2>0$ i.e.,
$\Gamma_2>\Gamma_0$, which means the criticality in this region possesses edge modes and is the topological non-trivial phase. Note that, the term $\epsilon_2$ is the gap term at criticality, which mimics the role of mass. As $\epsilon_2\rightarrow 0$ the decay length $\xi_+ \rightarrow \infty $ indicating the edge mode delocalize into the bulk and 
topological transition takes place at $MC_1$ i.e., at $\Gamma_2=\Gamma_0$. To visualize this phenomenon 
we write the bound state solution $\psi(x) \propto \begin{pmatrix}
\eta & 0
\end{pmatrix}^T (e^{-x/\xi_+}-e^{-x/\xi_-}),$
which distributes dominantly near the boundary and decay exponentially as $x\rightarrow\infty$, as shown in Fig.\ref{Dirac}(a). 

To identify the topological transition at $MC_2$ and the corresponding topological non-trivial phase one has to consider the swapping property of $k_0^{mc}$, 
which emerge as a result of intersection of critical lines at $MC_2$. In this case, after expanding around $k^{mc}_0$ and using the swapping property, the Dirac Hamiltonian can be obtained up to second order as 
\begin{equation}
\mathcal{H}(k) \approx (\epsilon_1-\epsilon_3k^2)\sigma_x +  (-i\epsilon_2 k) \sigma_y,
\end{equation}
where $\epsilon_1=2(\Gamma_0+\Gamma_2)$, $\epsilon_2=(\Gamma_0+3\Gamma_2)$ and $\epsilon_3=(5\Gamma_{2}+\Gamma_{0})/2$. 
With $\eta=\text{sign}(\epsilon_2/\epsilon_3)$,
the edge mode decay length $\xi_+ \approx -(|\epsilon_2|/\epsilon_1)$ is obtained using $\phi(x)\propto e^{x/\xi}$ and is positive
if $\epsilon_1<0$. 
Therefore, in this case, the gap term is $\epsilon_1$ which vanish at the multicritical point $MC_2$, i.e. at $\Gamma_2=-\Gamma_0$.
This implies that the criticality $\Gamma_2<-\Gamma_0$ is topological non-trivial 
phase and the topological transition occur at $MC_2$, i.e. $\Gamma_2=-\Gamma_0$ as a consequence of 
the delocalization of edge mode into the bulk as $\epsilon_1\rightarrow 0$. In this case the bound state solution $\psi(x) \propto \begin{pmatrix}
0 & \eta
\end{pmatrix}^T (e^{x/\xi_+}-e^{x/\xi_-})$,
distribute near the boundary and decay as  $x\rightarrow -\infty$ as shown in Fig.\ref{Dirac}(b).

\section{Winding number}\label{sec4}
The topological character of a gapped phase is quantified using topological invariant numbers~\cite{thouless1982quantized}. The quantized values of these invariant numbers represents the number of localized stable edge modes at each end of the open chain. For one dimensional systems winding number is a good invariant number which represents the winding of psuedospin vector in the Brillion zone~\cite{PhysRevLett.115.177204}.
\begin{figure}[t]
	\begin{minipage}[t]{.2\textwidth}
		\includegraphics[width=4.2cm,height=3.5cm]{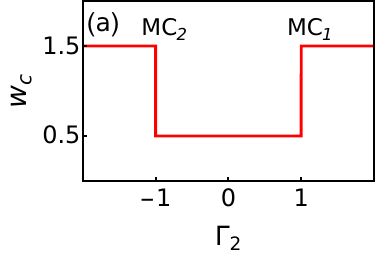}
	\end{minipage}
	\hspace{0.8cm} 
	\begin{minipage}[t]{.2\textwidth}
		\vspace{-3.8cm}
		\includegraphics[width=3.7cm,height=1.8cm]{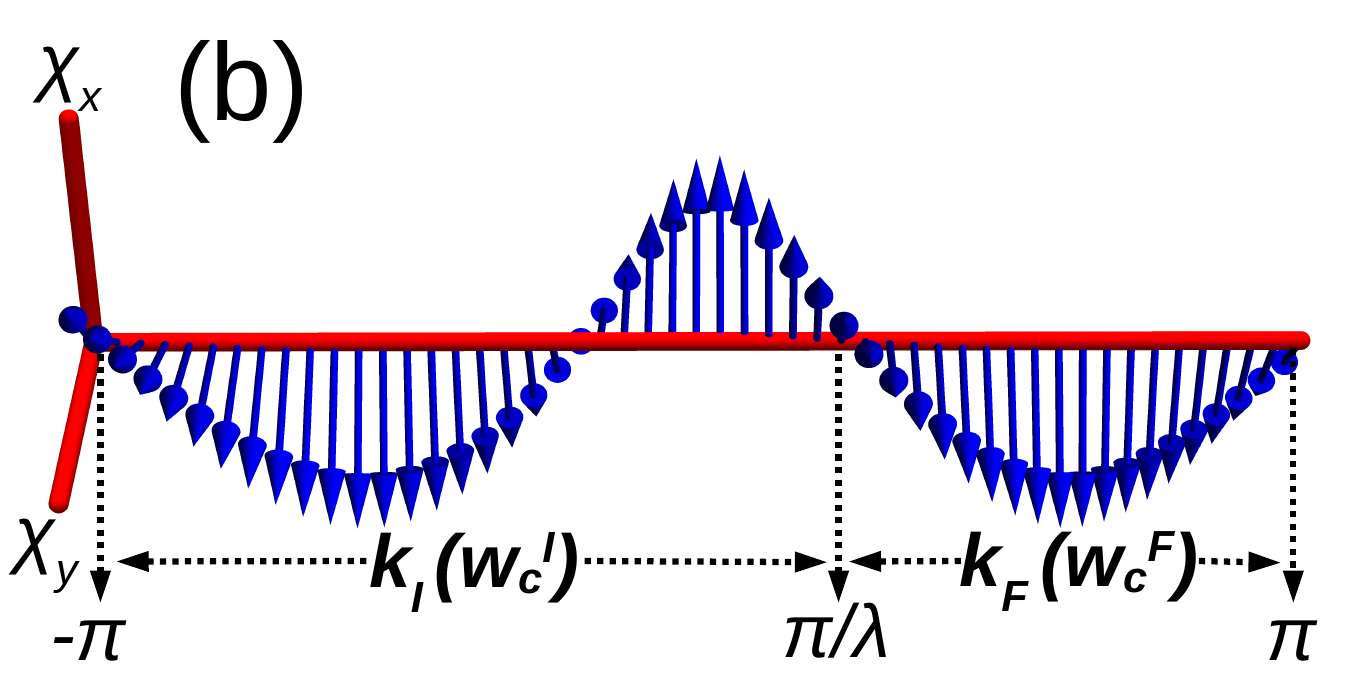}\\
		\vspace{0.2cm}
		\includegraphics[width=3.7cm,height=1.8cm]{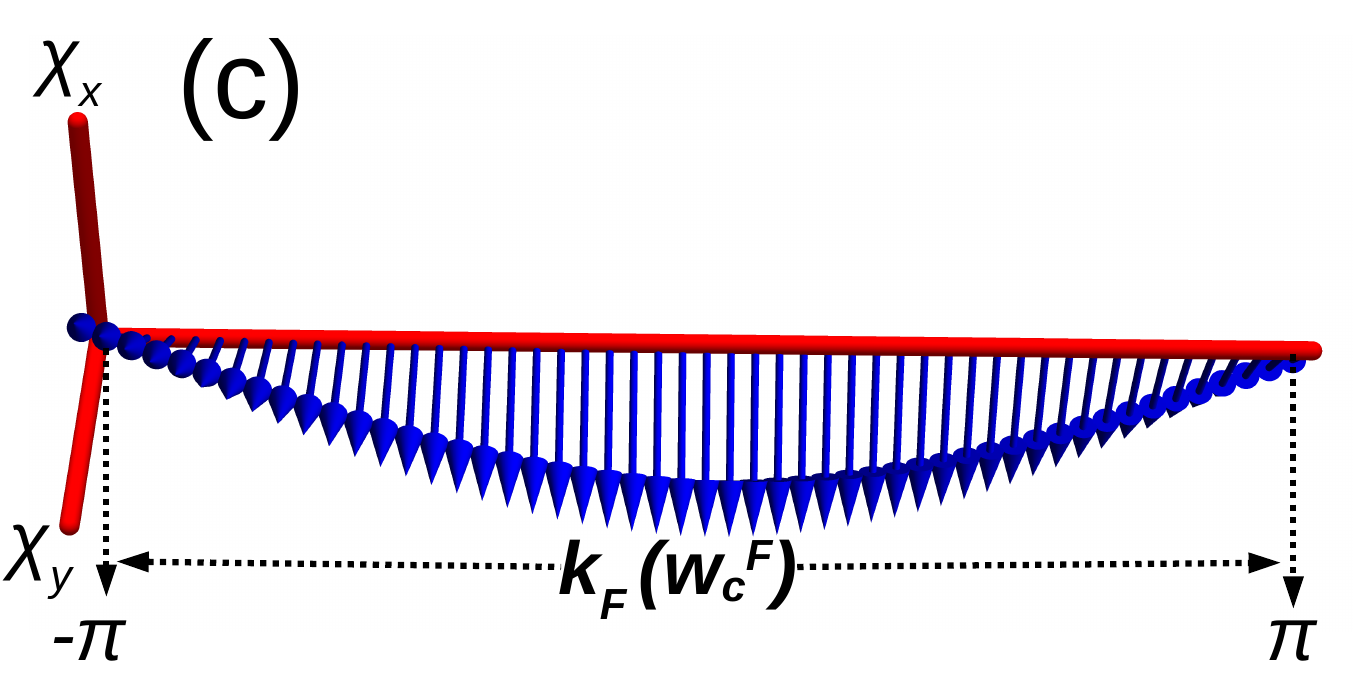}
	\end{minipage}
	\caption{\label{Winding-Fractional}\small Winding number at criticality (a) Fractional values of winding number ($w_c$). Trivial critical phases are identified with $w_c=0.5$ and non-trivial critical phases are identified with $w_c=1.5$. The topological transition 
		occurs at the multicritical points $MC_{1,2}$, i.e, $\Gamma_{2}=\pm \Gamma_{0}$ with $\Gamma_{0}=1$. (b) Winding of unit vector $\boldsymbol{\hat{\chi}}$ at non-trivial critical phases with both integer ($w_c^I$) and fractional parts ($w_c^F$). (c) Winding of $\boldsymbol{\hat{\chi}}$ at the trivial critical phase 
		with only fractional part ($w_c^F$).}
\end{figure}
Therefore,
the edge excitations of the gapped phases can be quantified in terms of winding number~\cite{hasan2010colloquium}
\begin{equation}
w=\frac{1}{2\pi}\oint\limits_{BZ}  F(k,\mathbf{\Gamma})  dk, \label{windings}
\end{equation}
where $ F(k,\mathbf{\Gamma}) = i \left\langle u_k\left| \partial_k\right| u_k\right\rangle$ is the 
Berry connection or curvature function of 
Bloch wavefunction $\psi_k(r)=u_k(r) e^{ikr}$.

In order to quantify the edge modes at criticality one has to define the winding number at criticality~\cite{verresen2020topology}. The conventional definition of winding number fails at criticalities.
This is due to the non-analyticity of the curvature function (integrand in Eq.\ref{windings}) at criticalities. This constraint is naturally avoided in the near-critical approach and allows one to calculate
the winding number in its usual integral 
form even at criticality.
\begin{equation}
w_c=\frac{1}{2\pi}\lim_{\delta \rightarrow 0}\oint\limits_{|k-k_0|>\delta} F(k,\mathbf{\Gamma}_c) dk \label{eq9}
\end{equation}
However, it yields fractional values, as shown in Fig.\ref{Winding-Fractional}(a),
which does not account correctly for the number of edge modes present at criticalities.

Alternatively, one can refer to the auxiliary space and differentiate between NN and NNN loops and consider only one among them which 
gives integer contribution and accounts for the edge modes at criticality \cite{rahul2021majorana}. However, this method is not efficient as the auxiliary space loops
gets complicated with the increasesing NN couplings \cite{PhysRevLett.115.177204}.
\begin{figure*}[t]
	\includegraphics[width=16.0cm,height=5.0cm]{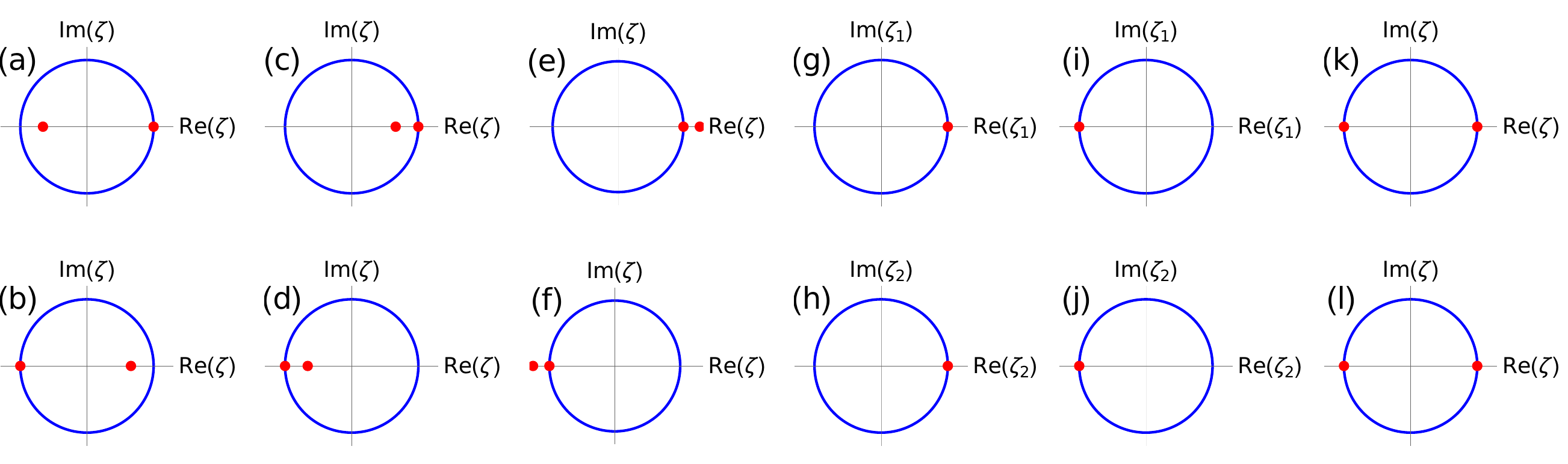}  
	\caption{\label{Winding-zeros} Zeros ($\zeta_{1,2}$) of the complex function in Eq.\ref{complex}, i.e. $\zeta_1=1$,  $\zeta_2=\Gamma_{0}/\Gamma_{2}$ for $\Gamma_{1}=-(\Gamma_{0}+\Gamma_{2})$ and  $\zeta_1=-1$,  $\zeta_2=-\Gamma_{0}/\Gamma_{2}$ for $\Gamma_{1}=(\Gamma_{0}+\Gamma_{2})$, represented as red dots. (a,b) Winding number $w=N_z-N_p=1$ at non-trivial critical phase with $\Gamma_{0}=1$ and $\Gamma_{2}=-1.5$. (c,d) Winding number $w=1$ at non-trivial critical phase with $\Gamma_{0}=1$ and $\Gamma_{2}=1.5$. (e,f) Winding number $w=0$ at trivial critical phase with $\Gamma_{0}=1$ and $\Gamma_{2}=0.5$. (g,h,i,j) Degenerate zeros on the unit circle at $MC_1$, for (g,h) the critical line $\Gamma_{1}=-(\Gamma_{0}+\Gamma_{2})$ and (i,j) the critical line $\Gamma_{1}=(\Gamma_{0}+\Gamma_{2})$. The figures (g,h,i,j) indicate the multicritical point $MC_1$ has multiplicity $m=2$ and dynamical exponent $z=2$~\cite{verresen2018topology}. (k,l) Zeros at $MC_2$. Both the zeros lie on the unit circle and are non-degenerate. The upper and lower panels are for $\Gamma_{1}=\mp(\Gamma_{0}+\Gamma_{2})$ respectively.}
\end{figure*}

\subsection{Winding number at criticality}\label{sec4a}
The fractional values at criticality imply the presence of fractional winding of unit vector $\boldsymbol{\hat{\chi}}=\boldsymbol{\chi}/|\boldsymbol{\chi}|$, in the auxiliary space over 
the Brillouin zone \cite{PhysRevLett.115.177204,verresen2020topology,rahul2021majorana}. For non-trivial critical phases, one can identify integer winding ($w^I_c$) of the unit vector
along with an extended fractional winding ($w^F_c$) in the Brillouin zone, as shown in Fig.\ref{Winding-Fractional}(b). For trivial criticalities, only fractional winding can be observed as in Fig.\ref{Winding-Fractional}(c). Based on this, we propose that the winding number at criticality should be approximated to only the integer values which effectively 
captures the number of edge modes at criticality. 

{\it Proposition:} \textit{Winding number at criticality ($w_c$), which acquires fractional values ($w_c=w^I_{c}+w^F_{c}$), can be 
	effectively approximated only to its integer part i.e. $w_c \approxeq w^I_{c}$, to quantify the number of edge modes present at criticality.}

The proposal roots in the fact that the momentum zones 
can be divided into integer and fractional windings as $-\pi< k_I<\pi/\lambda$ and
$\pi/\lambda<k_F<\pi$ respectively.
The cut-off $\lambda$ differentiates the momentum zones responsible for integer and fractional 
windings of the unit vector, as shown in Fig.\ref{Winding-Fractional}(b). Therefore, we write
\begin{align}
w_c&= \frac{1}{2\pi} \lim_{\delta \rightarrow 0}  \left(
\int\limits_{\substack{-\pi \\ |k-k_0|>\delta}}^{\pi/\lambda}  F(k,\mathbf{\Gamma}_c) dk 
+ \int\limits_{\substack{\pi/\lambda\\ |k-k_0|>\delta}}^{\pi}  F(k,\mathbf{\Gamma}_c) dk \right)\nonumber\\
&=w_c^I +w_c^F \nonumber \\
&\approxeq w_c^I \label{wind-integer}
\end{align}
The fractional winding can be found to have $w_c^F=1/2$ since the critical phases have one gap closing point in the Brillouin zone. The interger winding $w_c^I \in \mathbb{Z}$ counts the number of edge modes in the corresponding critical phase. The winding number in the non-trivial critical phases of the model can be found to have $w_c=1.5$, for which the corresponding $w^I_c=1$. Hence $w^I_c$ correctly accounts for one edge mode living at the criticalities which we also find from the solution of the Dirac equation. For the trivial critical phase $w_c=0.5$ and $w^I_c= 0$ implying no localized edge modes. The transition between the critical phases with $w^I_c=0$ and $w^I_c=1$ occur at the multicritical 
points. This clearly demonstrates the topological transition at criticality through multicritical points.

The proposal 
can be found viable for the critical systems that support non-trivial critical phases having the winding number $w_c>2$ and characterized with a single gap closing point.
The models with couplings beyond the second neighbor~\cite{kumar2022topological} support the non-trivial critical phases with winding numbers $w_c=2.5,3.5...$, etc.
In the case of $w_c=2.5$, the unit vector $\boldsymbol{\hat{\chi}}$ winds twice with an extended fractional winding in the Brillouin zone. The approximation to only the integer value yields $w^I_c=2$ which counts the two edge modes localized in the corresponding critical phase. Therefore, we expect that the proposition will be useful in characterizing the non-trivial critical phases with higher winding numbers and a single gap closing point. For more than one gap closing point in the Brillouin zone, such as the non-high symmetry points discussed in Ref.~\cite{kumar2022topological}, the proposition might need further modification.

\subsection{Winding number using zeros and poles}\label{sec4b}
The proposal and the integer winding number $w^I_{c}$ can be found consistent with the method used in Ref.\cite{verresen2018topology}, where the winding number is defined using number of zeros ($N_z$) and order of poles ($N_p$), $w=N_z-N_p$. The zeros and poles of a complex function is obtained by writing the fermionic creation and annihilation operators in terms of Majorana operators and followed by a Fourier transformation. With the substitution $e^{ik}=\zeta$, (where $\zeta$ is a complex number) where $e^{ik}$ goes around the unit circle in the complex plane as $k$ varies over the Brillouin zone, we get the complex function $f(\zeta)$ living on the unit circle in the
complex plane
\begin{equation}
f(\zeta) = \sum_{\mu=-\infty}^{\infty} t_{\mu} \zeta^{\mu}.
\end{equation}
For extended Kitaev model it reads $f(\zeta)=\sum_{\mu=0}^{2} t_{\mu} \zeta^{\mu}$ (with no poles) where $t_{0,1,2}$ are respectively $-\beta_0, \beta_1, \beta_2$ (parameters of Kitaev model in Eq.\ref{kitaev}). Using the mapping $2\beta_0=\Gamma_{0}$, $-2\beta_1=\Gamma_{1}$, and $-2\beta_2=\Gamma_{2}$ one can write the complex function for the generic model 
\begin{equation}
f(\zeta)=-\frac{\Gamma_{0}}{2}-\frac{\Gamma_{1}}{2}\zeta -\frac{\Gamma_{2}}{2} \zeta^2. \label{complex}
\end{equation}
The solutions are $\zeta_{1,2}=(\Gamma_{1}\pm \sqrt{\Gamma_{1}^2-4\Gamma_{0}\Gamma_{2}})/-2\Gamma_{2}$. To characterize the topological trivial and non-trivial critical phases we write the solution at criticalities. For $\Gamma_{1}=-(\Gamma_{0}+\Gamma_{2})$ we get $\zeta_1=1$, $\zeta_2=\Gamma_{0}/\Gamma_{2}$ and for
$\Gamma_{1}=(\Gamma_{0}+\Gamma_{2})$ we get $\zeta_1=-1$, $\zeta_2=-\Gamma_{0}/\Gamma_{2}$. 

It is evident that one of the zero lie on the unit circle since the system is critical and other zero falls inside (outside) the unit circle for topological non-trivial (trivial) critical phase, as shown in Fig.\ref{Winding-zeros}. Winding number is determined by the number of zeros falls inside the unit circle, whose value can be found consistent with $w_c^I$. For non-trivial critical phases as shown in Fig.\ref{Winding-zeros}(a,b,c,d) (where upper and lower panels represents $\Gamma_{1}=\mp(\Gamma_{0}+\Gamma_{2})$ respectively), $w=w_c^I=1$ and for trivial critical phases as shown in Fig.\ref{Winding-zeros}(e,f), $w=w_c^I=0$.

At $MC_1$, the zeros can be obtained to be degenerate (with multiplicity $m$) i.e. $\zeta_{1,2}=1$ in Fig.\ref{Winding-zeros}(g,h) and $\zeta_{1,2}=-1$ in Fig.\ref{Winding-zeros}(i,j) on the critical lines $\Gamma_{1}=-(\Gamma_{0}+\Gamma_{2})$ and $\Gamma_{1}=(\Gamma_{0}+\Gamma_{2})$ respectively. At $MC_2$, we get non-degenerate zeros with $\zeta_{1,2}=\pm 1$ for both the criticalities, as shown in Fig.\ref{Winding-zeros}(k,l).

\section{Curvature function}\label{sec5}
\begin{figure}[t]
	\begin{minipage}[t]{0.45\columnwidth}
		\includegraphics[width=3.8cm,height=2.3cm]{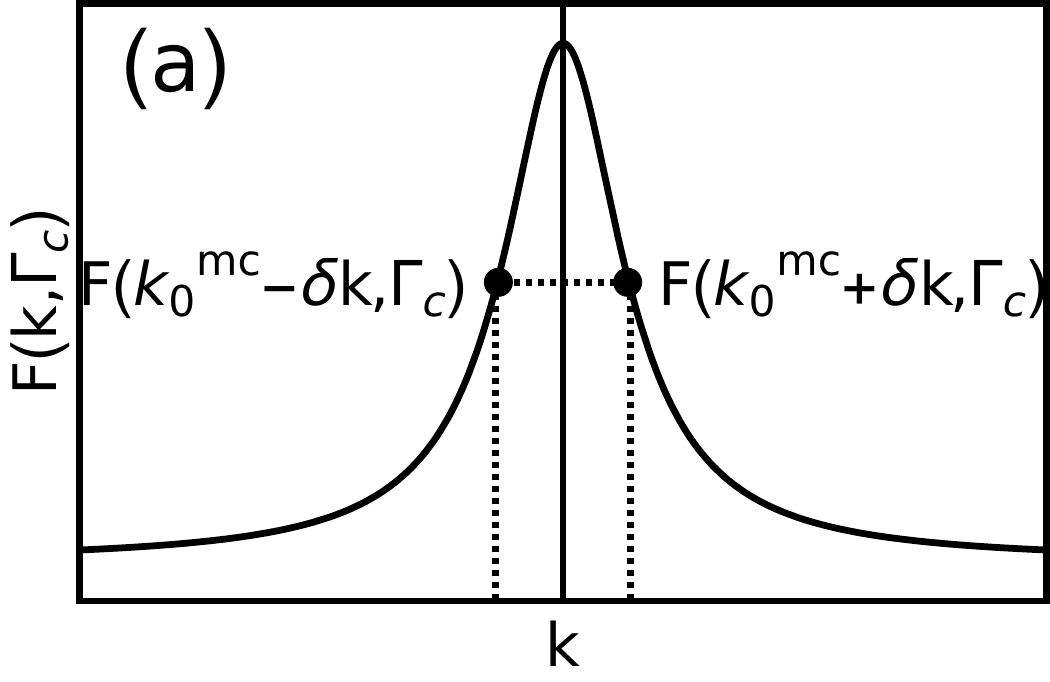}\\
		\vspace{0.3cm}
		\includegraphics[width=4cm,height=2.6cm]{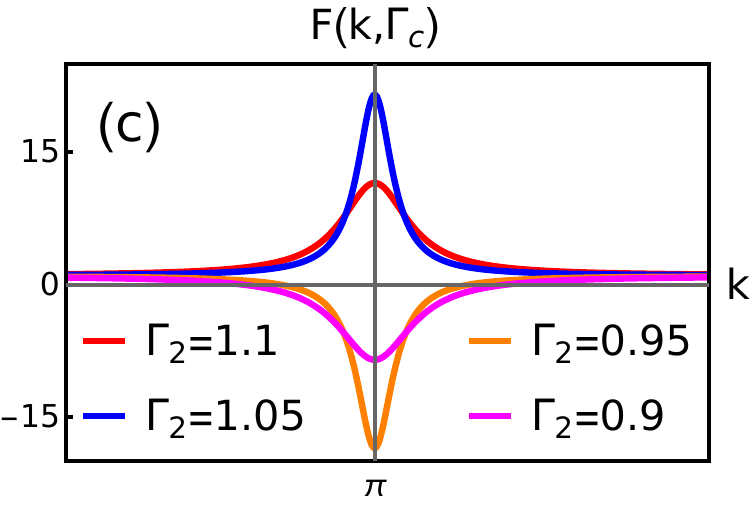} \\
		\vspace{0.3cm}
		\includegraphics[width=4cm,height=2.6cm]{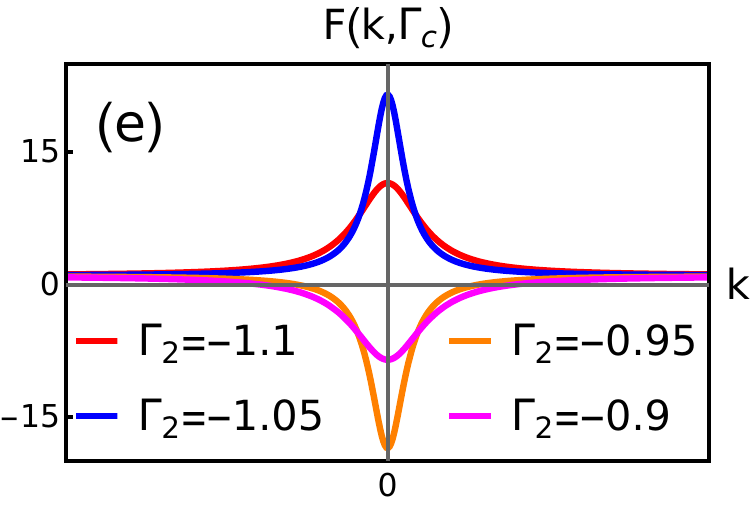} 
	\end{minipage}
\hfil
	\begin{minipage}[t]{0.45\columnwidth}
		\includegraphics[width=4cm,height=2.6cm]{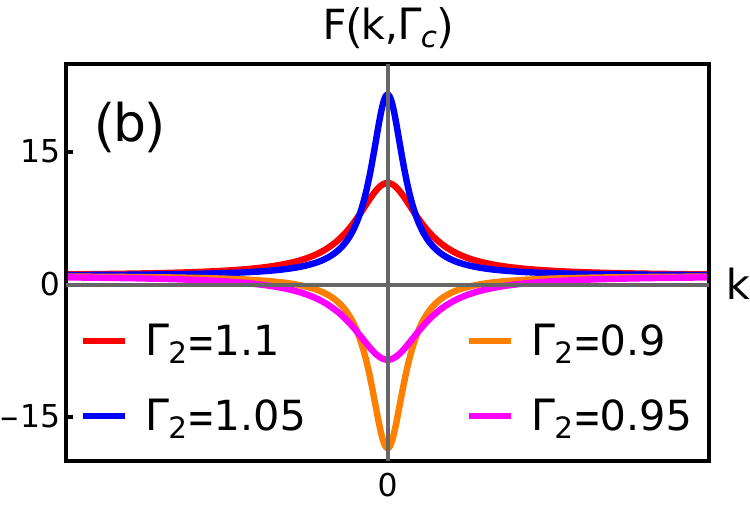}  \\
		\vspace{0.3cm}
		\includegraphics[width=4cm,height=2.6cm]{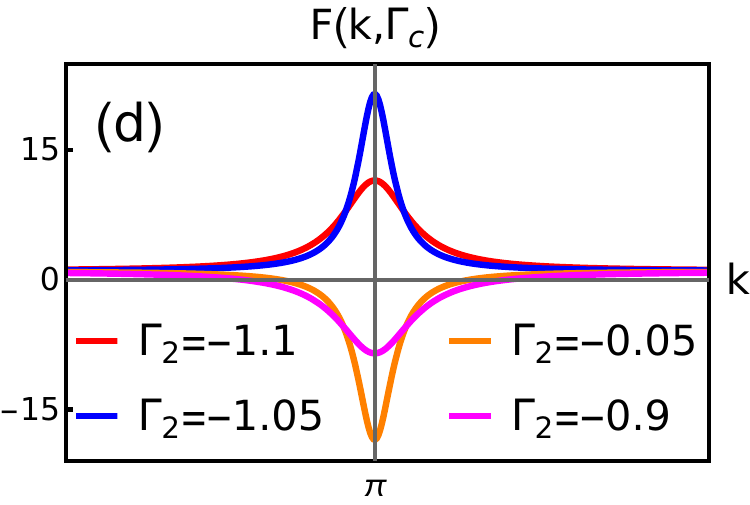} \\
		\vspace{0.6cm}
		\includegraphics[width=3.8cm,height=2.4cm]{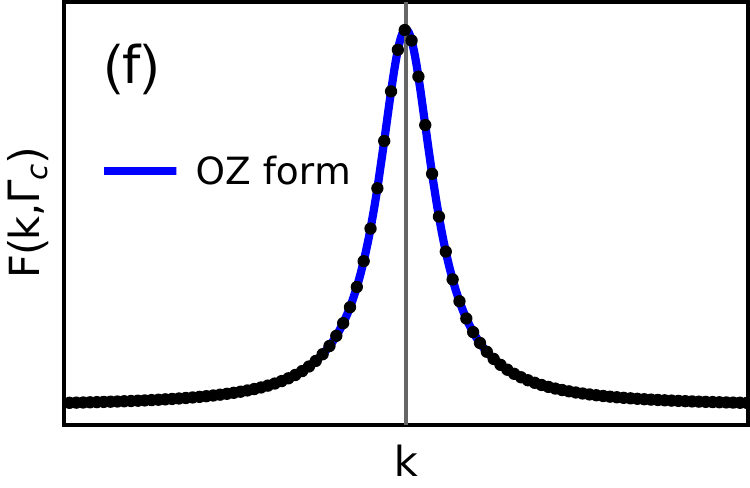}  
	\end{minipage}
	\caption{\label{curve-MC}Curvature function at criticality. (a) Illustration of symmetric nature of curvature function around $k_0^{mc}$ i.e. $F(k_0^{mc}+\delta k,\mathbf{\Gamma}_{c})=F(k_0^{mc}-\delta k,\mathbf{\Gamma}_{c})$. (b,c,d,e) Shows the diverging peaks of curvature function as the parameter $\Gamma_{2}$ tend towards the multicritical values $\Gamma_{2}=\pm \Gamma_0$ (with $\Gamma_0=1$). (b) For $MC_1$ at $\Gamma_1=-(\Gamma_0+\Gamma_2)$. (c) For $MC_1$ at $\Gamma_1=(\Gamma_0+\Gamma_2)$. (d) For $MC_2$ at $\Gamma_1=-(\Gamma_0+\Gamma_2)$. (e) For $MC_2$ at $\Gamma_1=(\Gamma_0+\Gamma_2)$. Flip in the sign of diverging peak is clearly observed as $\Gamma_{2}$ tuned across the multicritical points. The swapping of $k_0^{mc}$ for $MC_2$ is also evident from figures (d) and (e). (f) Shows the fitting of Ornstein-Zernike form in Eq.\ref{Lorenzian-crit} with the data points of curvature function at criticality.}
\end{figure}

Topological phase transition can be induced by changing the underlying topology of 
the system upon tuning the parameters $\mathbf{\Gamma}$ appropriately. 
The information of the topological property of the system is embedded in the 
curvature function 
$F(k,\mathbf{\Gamma})$ defined at momentum $k$ \cite{chen2016scaling,chen2016scalinginvariant,chen2018weakly,PhysRevB.95.075116,chen2019universality,molignini2018universal,molignini2020generating,abdulla2020curvature,malard2020scaling,molignini2020unifying,rufo2019multicritical,kartik2021topological}.
The topological quantum phase transition can be identified from the 
quantized jump of topological invariant number as the parameter tuned across
the critical point $\mathbf{\Gamma}_c$.
As the system approaches critical point to undergo 
topological phase transition i.e, $\mathbf{\Gamma}\rightarrow \mathbf{\Gamma}_{c}$, 
curvature function diverges at $k_0$, with the diverging curve satisfying 
$F(k_0+\delta k,\mathbf{\Gamma})=F(k_0-\delta k,\mathbf{\Gamma})$. The sign of the diverging peak 
flips across the critical point as
\begin{equation}
\lim_{\mathbf{\Gamma}\rightarrow \mathbf{\Gamma}_{c}^+}F(k_0,\mathbf{\Gamma})= -\lim_{\mathbf{\Gamma}\rightarrow \mathbf{\Gamma}_{c}^-}F(k_0,\mathbf{\Gamma})=\pm \infty. \label{div-curvature}
\end{equation}
Interestingly, even at criticality, the qualitative behavior of the curvature function remains the same with the fact that, 
now the critical point is a multicriticality which governs the topological transition between critical phases. As one tunes the parameters at criticality $\mathbf{\Gamma}_{c}$ towards a multicritical point $\mathbf{\Gamma}_{mc}$, the curvature function diverges at $k_0^{mc}$ with the symmetric nature $F(k_0^{mc}+\delta k,\mathbf{\Gamma}_{c})=F(k_0^{mc}-\delta k,\mathbf{\Gamma}_{c})$, as shown in Fig.\ref{curve-MC}(a).

Topological transition is signalled as the sign of the diverging peak flips if the parameters tuned across 
the multicritical point.
\begin{equation}
\lim_{\mathbf{\Gamma}_{c}\rightarrow \mathbf{\Gamma}_{mc}^+}F(k_0^{mc},\mathbf{\Gamma}_{c})= -\lim_{\mathbf{\Gamma}_{c}\rightarrow \mathbf{\Gamma}_{mc}^-}F(k_0^{mc},\mathbf{\Gamma}_{c})=\pm \infty. \label{div-curvature-crit}
\end{equation}
This is the characteristic feature of topological transition at criticality through both the multicritical points $MC_{1,2}$. The curvature function of the generic model at criticality can be written using the 
critical line relations of the parameters.  
The pseudo-spin vectors on the two critical lines, $\Gamma_1=\pm(\Gamma_0+ \Gamma_2)$, of the model  are 
$ \chi_{x} (k) =\Gamma_0 (1\pm \cos k) + \Gamma_2 (\cos 2k \pm \cos k),$ and $ \chi_{y} (k) = \Gamma_2 (\sin 2k \pm \sin k)\pm \Gamma_0 \sin k$.
This defines curvature function on the critical lines $F(k,\mathbf{\Gamma}_{c})=F(k,\mathbf{\Gamma}_{\Gamma_1=\pm(\Gamma_0+ \Gamma_2)})$,
\begin{align}
F(k,\mathbf{\Gamma}_{\Gamma_1=\pm(\Gamma_0 + \Gamma_2)}) &= \frac{\chi_x\partial_k \chi_y - \chi_y \partial_k \chi_x}{\chi_x^2+\chi_y^2} \nonumber \\
&=  \frac{\Gamma_{0}^2+3\Gamma_{2}^2\pm 4\Gamma_{0}\Gamma_{2} \cos k}{2(\Gamma_{0}^2+\Gamma_{2}^2\pm 2\Gamma_{0}\Gamma_{2} \cos k)} \label{curv-crit12}
\end{align}
The property in Eq.\ref{div-curvature-crit} can be observed to be obeyed by $F(k,\mathbf{\Gamma}_{\Gamma_1=\pm(\Gamma_0+ \Gamma_2)})$ as shown in the Fig.\ref{curve-MC}(b,c,d,e). They show the critical behavior of curvature function around 
the multicritical 
points $MC_{1,2}$, which distinguish between distinct critical phases. The peak of the curvature function tends to diverge as the parameters 
approach $MC_{1,2}$ from both sides at criticality. Both the criticalities 
exhibit the universal nature of curvature function around the multicritical points.

The scenario around $MC_1$ on the critical line $\Gamma_1=-(\Gamma_0+\Gamma_2)$ shows the divergence in curvature function at the $k_0^{mc}=0$, as shown in Fig.\ref{curve-MC}(b). As the parameter $\Gamma_2$ is tuned towards its multicritical value (i.e $MC_1$) on both sides, the diverging peak of curvature function increases leading to a complete divergence at $MC_1$ and flips sign as the
critical value is crossed. This signals the topological transition across $MC_1$ at criticality.
Similar behavior of curvature function can be observed around $MC_1$ on the critical line $\Gamma_1=(\Gamma_0+\Gamma_2)$, for 
which the divergence occurs at $k_0^{mc}=\pi$,
as shown in Fig.\ref{curve-MC}(c).

The nature of curvature function around $MC_2$ at both the criticalities share the same 
property of divergence and flipping of sign as shown in Fig.\ref{curve-MC}(d) and \ref{curve-MC}(e).
Note that, the $k_0^{mc}$ at which the diverging peak increases on approaching the multicritical value, is $k_0^{mc}=\pi$ instead of $k_0^{mc}=0$ for $\Gamma_1=-(\Gamma_0+\Gamma_2)$ (and $k_0^{mc}=0$ instead of $k_0^{mc}=\pi$ for $\Gamma_1=(\Gamma_0+\Gamma_2)$). 
This swapping of $k_0^{mc}$ occur as a consequence of the intersection of critical lines. Typically the multicritical point $MC_2$ is the same point for both the critical lines $\Gamma_1=\pm(\Gamma_0+\Gamma_2)$ in parameter space. These critical lines intersect each other at
$MC_2$, which results in the swapping of respective $k_0^{mc}$ values.

\subsection{Critical exponents}\label{expo}
The condition in Eq.\ref{div-curvature} for curvature function allows one to choose the proper gauge for which $F(k,\mathbf{\Gamma})$ 
can be written in Ornstein-Zernike form around the $k_0$ \cite{chen2016scaling},
\begin{equation}
F(k_0+\delta k,\mathbf{\Gamma})=\frac{F(k_0,\mathbf{\Gamma})}{1+\xi^2\delta k^2}, \label{Lorenzian}
\end{equation}
where $\delta k$ is small deviation from $k_0$, $F(k_0,\mathbf{\Gamma})$ is the 
height of the peak and $\xi$ is characteristic 
length scale or the width of the peak. As we approach critical point, one can also find the divergence in the 
characteristic length $\xi$ 
along with the
curvature function.
The divergences in both $F(k_0,\mathbf{\Gamma})$ and $\xi$ give rise to 
the critical exponents
\begin{equation}
F(k_0,\mathbf{\Gamma}) \propto |\mathbf{\Gamma}-\mathbf{\Gamma}_c|^{-\gamma},\;\;\;\;\;\; \xi \propto |\mathbf{\Gamma}-\mathbf{\Gamma}_c|^{-\nu}, \label{critical-exponents}
\end{equation}
where $\gamma$ and $\nu$ are the critical exponents which define the universality 
class of the  
undergoing topological phase transition. These exponents obeys a 
scaling law, imposed by the conservation of topological invariant, which reads 
$\gamma=\nu$ for 1D systems \cite{PhysRevB.95.075116}.

\begin{figure}[t]
	\includegraphics[width=4.2cm,height=3.0cm]{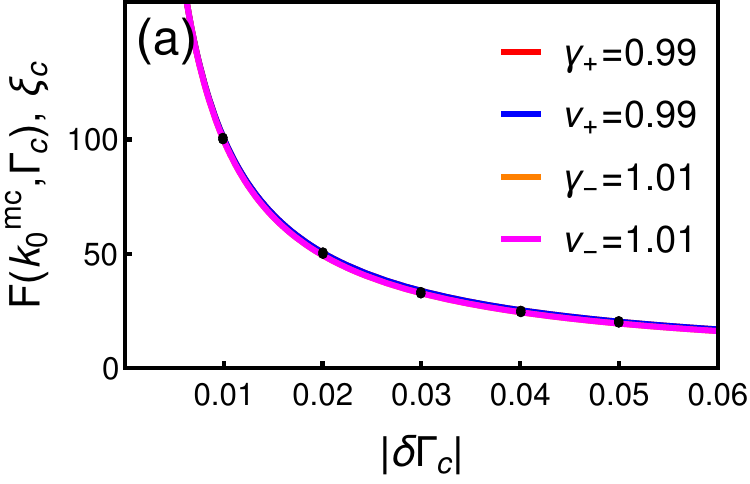} 
	\includegraphics[width=4.2cm,height=3.0cm]{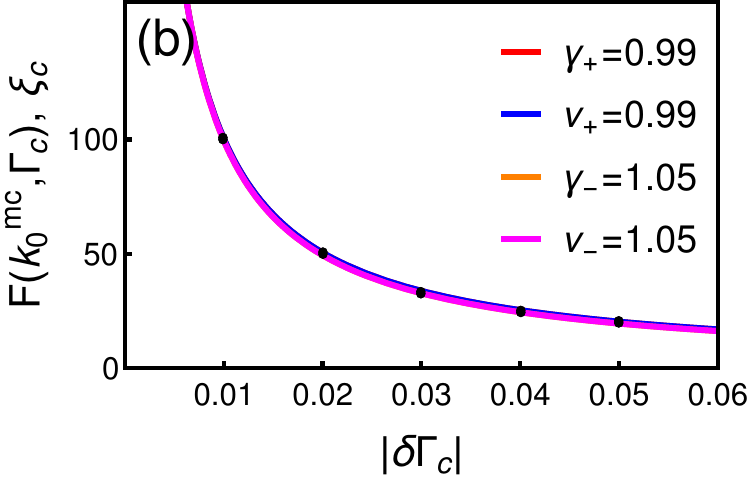} 
	\includegraphics[width=4.2cm,height=3.0cm]{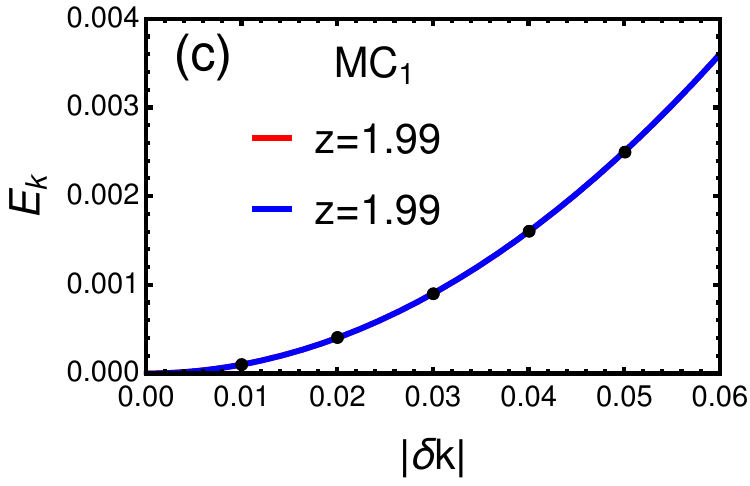} 
	\includegraphics[width=4.2cm,height=3.0cm]{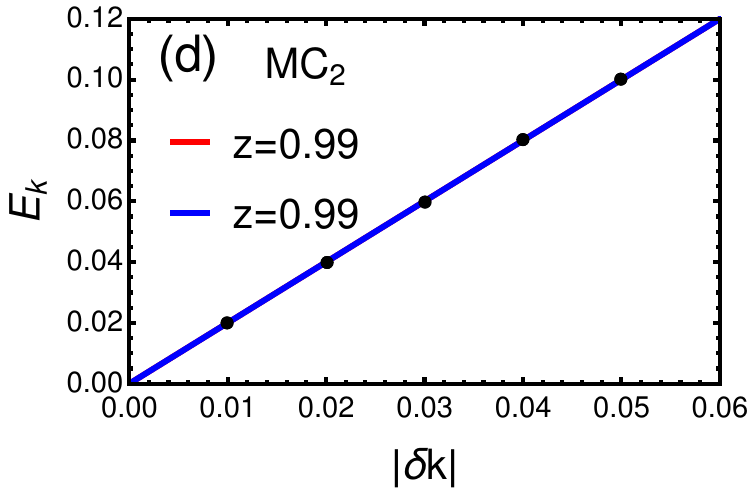} 
	\caption{\label{exponents}Critical exponents. Figures (a) and (b) represents exponents of curvature function ($\gamma$ and $\nu$) for $MC_1$ and $MC_2$ respectively.
		The notation $\gamma_{+/-}$ and $\nu_{+/-}$ represents the exponents on approaching the
		multicritical points from either sides. Dynamical exponent for, (c) $MC_1$: represents quadratic
		dispersion and (d) $MC_2$: represents linear dispersion. Red and Blue in (c,d) corresponds to the criticalities $\Gamma_{1}=\mp(\Gamma_{0}+\Gamma_{2})$ respectively.}
\end{figure}

Surprisingly, these scaling behavior of curvature function also appear at multicriticality by approaching it along the critical lines.
Approaching 
multicritical points $MC_{1,2}$, curvature function acquires Ornstein-Zernike form around $k_0^{mc}$. 
\begin{equation}
F(k_0^{mc}+\delta k,\mathbf{\Gamma}_{c})=\frac{F(k_0^{mc},\mathbf{\Gamma}_{c})}{1+\xi_{c}^2\delta k^2}, \label{Lorenzian-crit}
\end{equation}
where $\delta k=|k-k_0^{mc}|$, $\xi_{c}$ is the characteristic length scale at criticality and it represents the width 
of the curvature function that develops around $k_0^{mc}$ as the parameters $\mathbf{\Gamma}_{c}\rightarrow\mathbf{\Gamma}_{mc}$. The critical behavior of curvature function around the multicritical points $MC_{1,2}$ can be captured
by the same exponents $\gamma$ and $\nu$ defined by 
\begin{equation}
F(k^{mc}_0,\mathbf{\Gamma}_{c}) \propto |\mathbf{\Gamma}_{c}-\mathbf{\Gamma}_{mc}|^{-\gamma},\;\;\; \xi_{c} \propto |\mathbf{\Gamma}_{c}-\mathbf{\Gamma}_{mc}|^{-\nu}. \label{critical-exponents-crit}
\end{equation} 
One can calculate these critical exponents and quantify the scaling properties, numerically, through fitting the diverging peak of curvature function with the Ornstein-Zernike form in Eq.\ref{Lorenzian-crit}, as shown in Fig.\ref{curve-MC}(f). The data points collected for $F(k^{mc}_0,\mathbf{\Gamma}_{c})$ and $\xi_{c}$ can then be fitted again with the equation of the form in Eq.\ref{critical-exponents-crit}, to extract the exponents $\gamma$ and $\nu$ at the multicritical points. Fig.\ref{exponents}(a) and (b) shows the acquired values of exponents for $MC_1$ and $MC_2$ respectively, on approaching from either sides.
The critical exponents are found to be, $\gamma_{+/-}=\gamma\approx 1$ and 
$\nu_{+/-}=\nu\approx 1$ for both multicritical points $MC_{1,2}$, 
where $\gamma_{+(-)}$ and 
$\nu_{+(-)}$ represents the scaling behavior of curvature function with positive (negative) peaks around 
the multicritical points on both the criticalities.

The exponents can also be estimated analytically by writing the curvature function in Ornstein-Zernike form (see Appendix.\ref{expo-nume} for details). It yield the same values of critical exponents for both $MC_{1,2}$.
The exponents calculated 
obeys certain scaling laws and defines universality class of the multicriticalities.
For topological transition occurring through both the multicritical point $MC_{1,2}$ the exponents are found to have $\gamma=\nu=1$ both numerically and analytically. The scaling law $\gamma=\nu$ for 1D systems \cite{PhysRevB.95.075116} is thus true for the critical behavior of the multicritical points governing the topological transition at criticality. 

In addition, the dynamical exponent $z$ dictates the nature of the spectra near the gap closing momenta $k_0^{mc}$, i.e. $E_k\propto k^z$~\cite{verresen2018topology}.
It can be calculated numerically 
as shown in the Fig.\ref{exponents}(c) and (d), where the data points around gap closing momenta $k_0^{mc}$ at the 
multicritical points $MC_{1,2}$ are shown. The spectra is quadratic at $MC_1$ and linear at $MC_2$. The quadratic spectra results in the dynamical critical exponent $z\approx 2$, whilst for linear spectra $z\approx1$. This behavior is true for both the criticalities. Therefore, the multicriticalities with both  $z = 1$ and $z = 2$ favour the topological transition at criticality. 

The universality class for the topological transition at criticality through both $MC_{1,2}$ can now be obtained using the set of three critical
exponents $(\gamma,\nu,z)$, which captures the scaling behavior around the multicritical points with distinct nature. The universality class of the multicriticality at $MC_1$ is $(1,1,2)$ and for $MC_2$ it reads $(1,1,1)$. Therefore, it is clear that the topological transition at quantum criticality occurs through two distinct multicriticalities which belongs to different universality classes. 

\subsection{Scaling theory}\label{sec5b}
Based on the divergence of the curvature function, a scaling 
theory has been developed \cite{chen2016scaling,chen2016scalinginvariant,chen2018weakly,chen2019universality,molignini2018universal,molignini2020generating,abdulla2020curvature,chen2016scaling,malard2020scaling,molignini2020unifying,kumar2021multi}. This is achieved by the 
deviation reduction mechanism where the deviation of the curvature function from its
fixed point configuration can be reduced gradually. In the curvature function $F(k,\mathbf{\Gamma})$, 
for a given $\mathbf{\Gamma}$ in the parameter space, one can find a new 
$\mathbf{\Gamma^{\prime}}$ which satisfies
\begin{equation}
F(k_0,\mathbf{\Gamma^{\prime}})=F(k_0+\delta k,\mathbf{\Gamma}), \label{scaling}
\end{equation}
where $\delta k$ is small deviation away from the $k_0$, satisfying $F(k_0+\delta k,\mathbf{\Gamma})=F(k_0-\delta k,\mathbf{\Gamma})$. 
As a consequence of the same 
topology of the system at $\mathbf{\Gamma}$ and at fixed point $\mathbf{\Gamma}_f$, the curvature function can be written as $F(k,\mathbf{\Gamma})=F_f(k,\mathbf{\Gamma}_f)+F_d(k,\mathbf{\Gamma}_d)$, 
where  $F_f(k,\mathbf{\Gamma}_f)$ is curvature function at fixed point 
and $F_d(k,\mathbf{\Gamma}_d)$ is deviation from the fixed point. The scaling procedure drives the deviation part of curvature function $|F_d(k_0,\mathbf{\Gamma}_{d})|\rightarrow 0$.
The fixed point configuration is invariant under the scaling operation 
i.e., $F_f(k_0,\mathbf{\Gamma}_f)=F_f(k_0+\delta k,\mathbf{\Gamma}_f)$.

Performing the scaling procedure in Eq.\ref{scaling} iteratively and solving $\mathbf{\Gamma}$ for every deviation $\delta k$, one can
obtain a renormalization group (RG)  
equation for the coupling parameters. Expanding Eq.\ref{scaling} in leading order and 
writing $\mathbf{\Gamma^{\prime}}-\mathbf{\Gamma} = d\mathbf{\Gamma}$ and $(\delta k)^2=dl$, one can obtain a generic RG equation
\begin{equation}
\frac{d\mathbf{\Gamma}}{dl}= \frac{1}{2} \frac{\partial_k^2 F(k,\mathbf{\Gamma})|_{k=k_0}}{\partial_{\mathbf{\Gamma}}F(k_0,\mathbf{\Gamma})}. \label{CRG_eq}
\end{equation}
Since the curvature function diverges
at $\mathbf{\Gamma}_c$, the scaling procedure gradually drives the system away from $\mathbf{\Gamma}_c$
towards $\mathbf{\Gamma}_f$ without changing the topological invariant. Thus,
eventually, 
the RG flow distinguishes between distinct gapped phases and correctly captures the topological phase transitions
between the gapped phases in the system.

In order to capture the topological transition at criticality one can modify the same scaling scheme to incorporate the multicriticality. This is possible since the qualitative behavior of the curvature function defined at criticality exhibits the same diverging nature near multicritical points with the property $F(k_0^{mc},\mathbf{\Gamma^{\prime}}_{c})=F(k_0^{mc}+\delta k,\mathbf{\Gamma}_{c})$ (here $\delta k$ is small deviation from $k_0^{mc}$). As the parameters at criticality $\mathbf{\Gamma}_{c} \rightarrow \mathbf{\Gamma}_{mc}$, the topology of the critical phase changes implying a topological transition at multicritical point.

Based on the divergence of the curvature function at criticality, the scaling 
theory can be achieved by performing the 
deviation reduction mechanism at criticality.
As a consequence of the same 
topology of the system at $\mathbf{\Gamma}_c$ and at fixed point $\mathbf{\Gamma}_c^f$, the curvature function can be written as $F(k,\mathbf{\Gamma}_c)=F_f(k,\mathbf{\Gamma}_c^f)+F_d(k,\mathbf{\Gamma}_c^d)$,
where  $F_f(k,\mathbf{\Gamma}_c^f)$ is the curvature function at fixed point 
and $F_d(k,\mathbf{\Gamma}_c^d)$ is deviation from the fixed point.
For a given $\mathbf{\Gamma}_c$, one can find a new 
$\mathbf{\Gamma^{\prime}}_c$ which satisfies $F(k_0^{mc},\mathbf{\Gamma^{\prime}}_c)=F(k_0^{mc}+\delta k,\mathbf{\Gamma}_c)$.
Iteratively performing this scaling procedure and solving $\mathbf{\Gamma}_c$ for every $\delta k$, deviation of curvature function decreases and eventually $ F(k,\mathbf{\Gamma}_c) \rightarrow F_f(k,\mathbf{\Gamma}_c^f)$.
\begin{figure}[t]
	\centering
	\includegraphics[width=4.0cm,height=3.5cm]{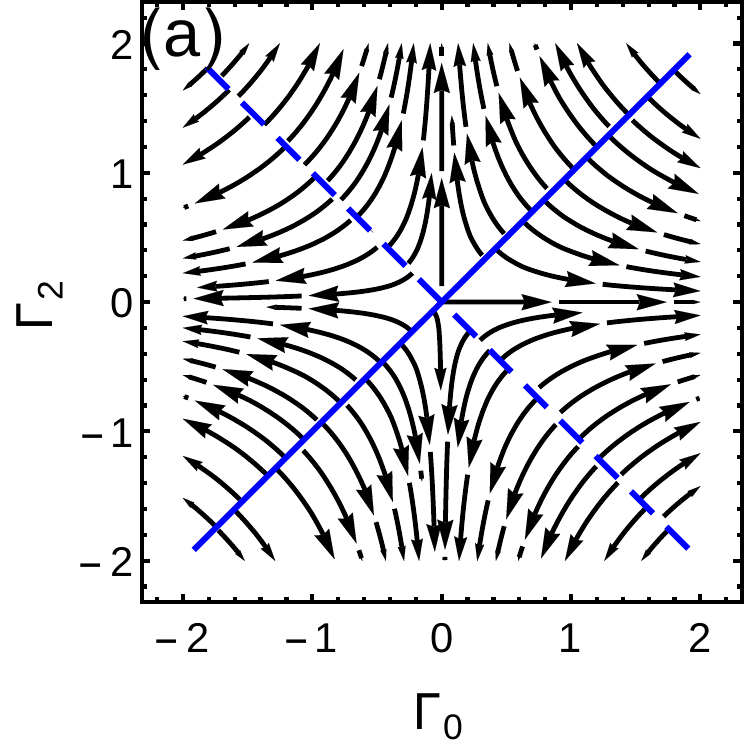}\hspace{0.25cm}  
	\includegraphics[width=4.0cm,height=3.5cm]{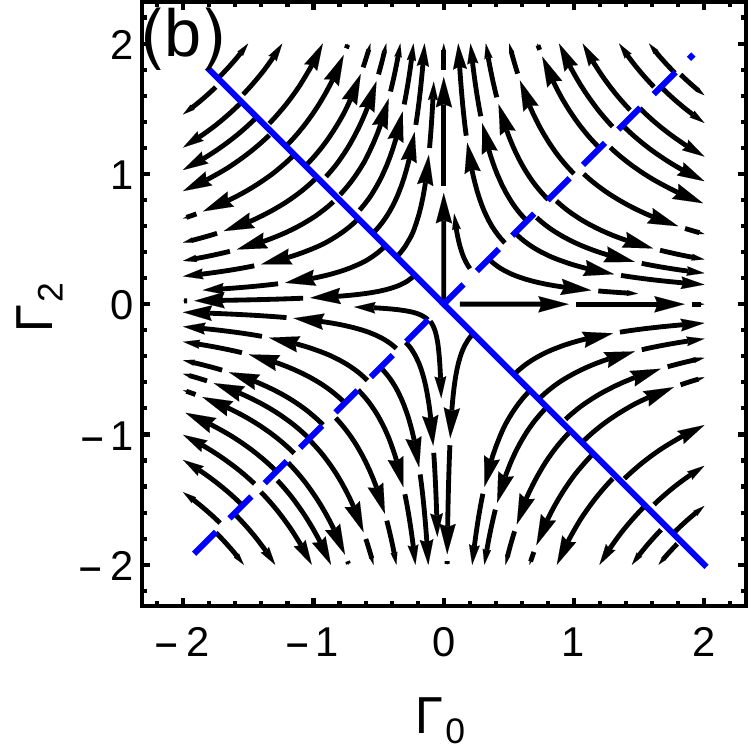}  
	\caption{\small \label{CRG-Curve}RG flow diagrams. The multicritical and fixed lines are represented as solid and dashed lines respectively. (a) For $MC_1$,  where $MC_2$ appears as unstable fixed line. (b) For $MC_2$, where $MC_1$ appears as unstable fixed line. The RG flow lines clearly demonstrates the topological transition at criticality.}
\end{figure}

One can obtain a renormalization group (RG) equation for the coupling parameters using the scaling parameter $\delta k^2=dl$ and $\mathbf{\Gamma^{\prime}}_c-\mathbf{\Gamma}_c = d\mathbf{\Gamma}_c$ as
\begin{equation}
\frac{d\mathbf{\Gamma}_{c}}{dl}= \frac{1}{2} \frac{\partial_k^2 F(k,\mathbf{\Gamma}_{c})|_{k=k_0^{mc}}}{\partial_{\mathbf{\Gamma}_{c}}F(k_0^{mc},\mathbf{\Gamma}_{c})}. \label{CRG_eq_crit}
\end{equation}
The distinct critical phases with different topological characters can be distinguished from the RG flow of Eq.\ref{CRG_eq_crit}. The
multicritical points and fixed points are then easily captured by analyzing the RG flow lines.
\begin{align}
\text{Multicritical point:}& \hspace{0.2cm} \left| \frac{d\mathbf{\Gamma}_{c}}{dl}\right| \rightarrow \infty, \text{flow directs away}.\nonumber\\
\text{Stable fixed point:}& \left| \frac{d\mathbf{\Gamma}_{c}}{dl}\right| \rightarrow 0, \text{flow directs into}.\nonumber\\
\text{Unstable fixed point:}& \left| \frac{d\mathbf{\Gamma}_{c}}{dl}\right| \rightarrow 0, \text{flow directs away}.
\label{crit-fixed}
\end{align}
Performing the RG scheme to the model at criticality, we obtain the RG equations for $MC_1$
as
\begin{equation}
\dfrac{d\Gamma_0}{dl} = \frac{\Gamma_0(\Gamma_0+\Gamma_2)}{2(\Gamma_0-\Gamma_2)} \hspace{0.3cm} \text{and} \hspace{0.3cm} \dfrac{d\Gamma_2}{dl} = -\frac{\Gamma_2(\Gamma_0+\Gamma_2)}{2(\Gamma_0-\Gamma_2)}
\end{equation}
Both the critical lines $\Gamma_1=\pm(\Gamma_0+\Gamma_2)$,
yield the same RG equations. 
The multicritical point $MC_1$ is manifested as a line $\Gamma_0=\Gamma_2$ with all flow lines flowing away, as shown in Fig.\ref{CRG-Curve}(a). 
The condition in Eq.\ref{crit-fixed} for multicritical points is satisfied as the flow rate diverges at 
$MC_1$, which also indicate that it is the
topological phase transition point between critical phases.
Surprisingly, $\Gamma_0=-\Gamma_2$ ($MC_2$) is obtained as a line of unstable fixed points at which flow rate vanishes with all the flow lines are flowing away.

In order to realize the topological transition at criticality through $MC_2$ one has to consider the swapping of $k_0^{mc}$. The RG equation 
for the critical line $\Gamma_1=(\Gamma_0+\Gamma_2)$, has to be derived with $k_0^{mc}=0$ and vice versa. This procedure yield the RG equations of the form
\begin{equation}
\dfrac{d\Gamma_0}{dl} = \frac{\Gamma_0(\Gamma_0-\Gamma_2)}{2(\Gamma_0+\Gamma_2)} \hspace{0.5cm} \text{and} \hspace{0.5cm} \dfrac{d\Gamma_2}{dl} = -\frac{\Gamma_2(\Gamma_0-\Gamma_2)}{2(\Gamma_0+\Gamma_2)}
\end{equation}
In this case, $\Gamma_0=-\Gamma_2$ ($MC_2$) is obtained to be 
the topological transition point between critical phases, with the diverging flow rate and flow lines directing away, as shown in Fig.\ref{CRG-Curve}(b). The unstable fixed point appear at $\Gamma_0=\Gamma_2$ ($MC_1$) with vanishing flow rate and flow lines flowing away. 


\subsection{Wannier state correlation function}\label{sec5c}
Along with the RG scheme, a correlation function in terms of Wannier-state representation
is proposed 
to characterize the topological phase transition \cite{PhysRevB.95.075116}. This quantity may be measured in higher dimensions \cite{chen2019universality,PhysRevLett.110.165304,duca2015aharonov}.
It is the filled-band contribution
to the charge-polarization correlation between Wannier
states at different positions, and can be obtained after 
the Fourier transform of the curvature function.
For the two-band model considered here with only the lower band occupied  
the Wannier state at a
distance $R$,
\begin{equation}
\ket{R} = \int dk e^{ik(\hat{r}-R)}\ket{u_k}
\end{equation}
with position operator $\hat{r}$, defines Wannier state correlation function as the overlap of the states
$\ket{0}$ at the origin and at a distance $\ket{R}$, as \cite{PhysRevB.95.075116}
\begin{equation}
\lambda_R= \left\langle   R|\hat{r}|0\right\rangle = \int dk e^{ikR} \left\langle u_k |i\partial_k|u_k \right\rangle 
\end{equation} 
Meanwhile,
the substitution of the Ornstein-Zernike form of curvature function (Eq.\ref{Lorenzian}) yields the Wannier state correlation function $\lambda_R$, to be
\begin{equation}
\lambda_R = \int\limits \frac{dk}{2\pi} e^{ikR}F(k,\mathbf{\Gamma}) = e^{ik_0R} \frac{F(k_0,\mathbf{\Gamma})}{2\xi}e^{-\frac{R}{\xi}}. \label{correlation}
\end{equation}
where $\xi$ can be 
treated as correlation length of topological phase transition. The correlation function $\lambda_R$ decays exponentially on either sides of the critical point. The decay gets slower as the parameter is tuned towards criticality. 
\begin{figure}[t]
	\centering
	\includegraphics[width=4.2cm,height=3.2cm]{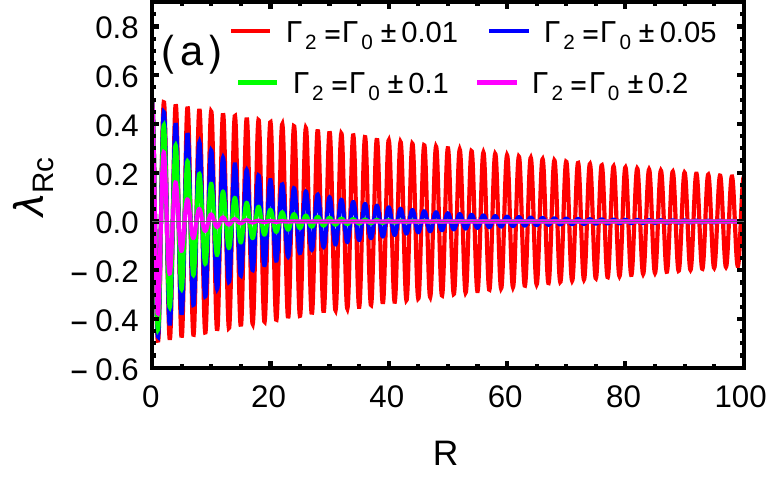}  
	\includegraphics[width=4.2cm,height=3.2cm]{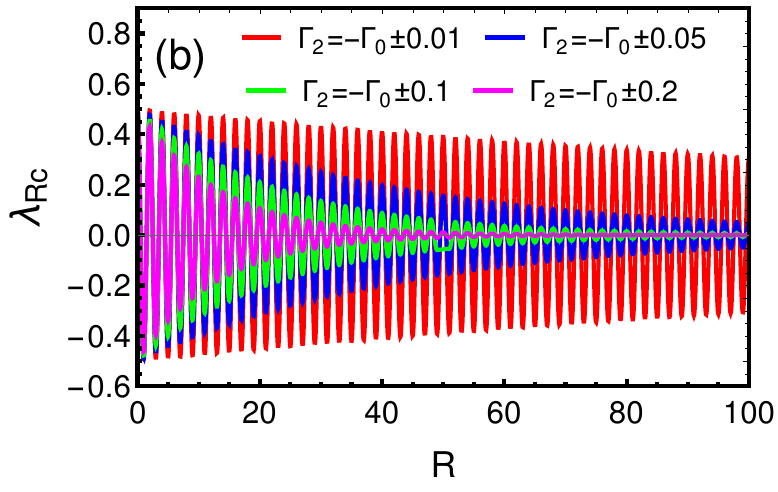}  
	\caption{\label{correl-wannier} Wannier state correlation function at criticality. (a) For $MC_1$. (b) For $MC_2$. Approaching the multicritical points $\Gamma_{2}=\pm\Gamma_{0}$ (with $\Gamma_{0}=1$), the decay in the correlation function gets slower on either sides of $MC_{1,2}$.}
\end{figure}

Surprisingly, this notion of correlation function holds true even at criticality and identify
the unique topological phase transition at criticality. The behavior of correlation
function evidently show that the topological phase transition occurs at the multicritical points $MC_{1,2}$ at both the criticalities. 
The Wannier state correlation function can be calculated at criticality as
\begin{equation}
\lambda_{Rc}=e^{ik^{mc}_0R} \frac{F(k_0^{mc},\mathbf{\Gamma}_{c})}{2 \xi_{c}} e^{-R/\xi_{c}}.
\end{equation}
where $\xi_{c}= F(k_0^{mc},\mathbf{\Gamma}_{c})= (\Gamma_{0}-3\Gamma_{2})/2(\Gamma_{2}-\Gamma_{0})$ for $MC_1$. The correlation function decays faster away from the  
the multicritical point $MC_1$ and the decay slow down as one approaches $MC_1$  with the correlation length $\xi_{c} \rightarrow \infty$, as shown in Fig.\ref{correl-wannier}(a). Both the criticalities shows same behavior
of correlation function near this multicritical point on both sides indicating that the multicriticality is indeed a topological 
phase transition point at criticality. Note that, the only difference between the criticalities for $k_0=0$ and $\pi$ is the oscillatory behavior of
$\lambda_{Rc}$ originating from the term $e^{ik_0R}$.

To obtain the critical nature of $MC_2$ one has to consider the swapping of $k^{mc}_0$ (Eq.\ref{swap-1} and Eq.\ref{swap-2}), which yields
$\xi_{c}=F(k_0^{mc},\mathbf{\Gamma}_{c})= (\Gamma_{0}+3\Gamma_{2})/2(\Gamma_{0}+\Gamma_{2})$. This captures the critical nature of $MC_2$, where the decay gets slower 
as one approaches this point from both the sides, as shown in Fig.\ref{correl-wannier}(b). Therefore, the behavior of the correlation function evidently shows that the topological phase transition occurs at the multicritical points. For both the criticalities, the correlation length $\xi_{c}$ coinsides with the decay length of the edge modes at criticality studied earlier.

\begin{figure}[t]
		\includegraphics[width=4.2cm,height=3.2cm]{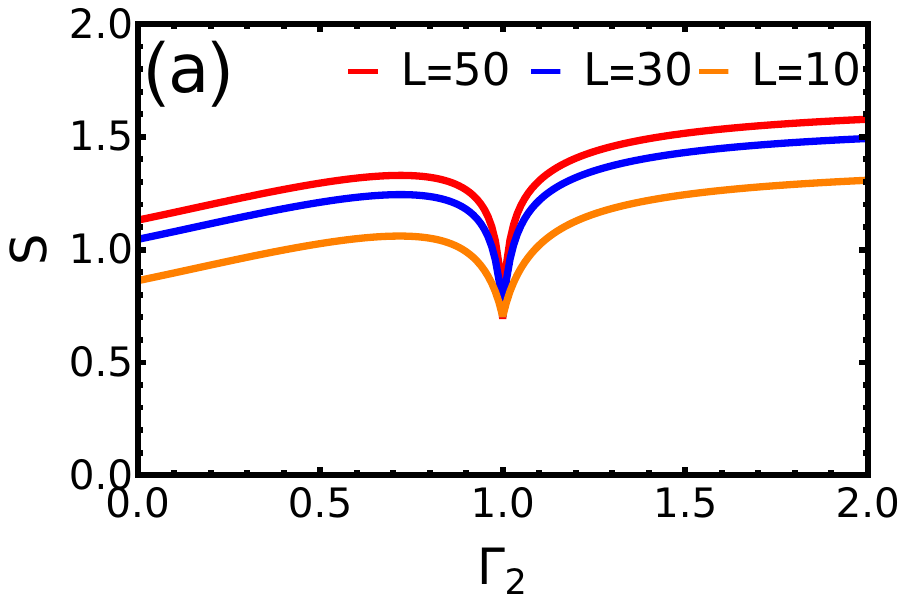} 
		\includegraphics[width=4.2cm,height=3.2cm]{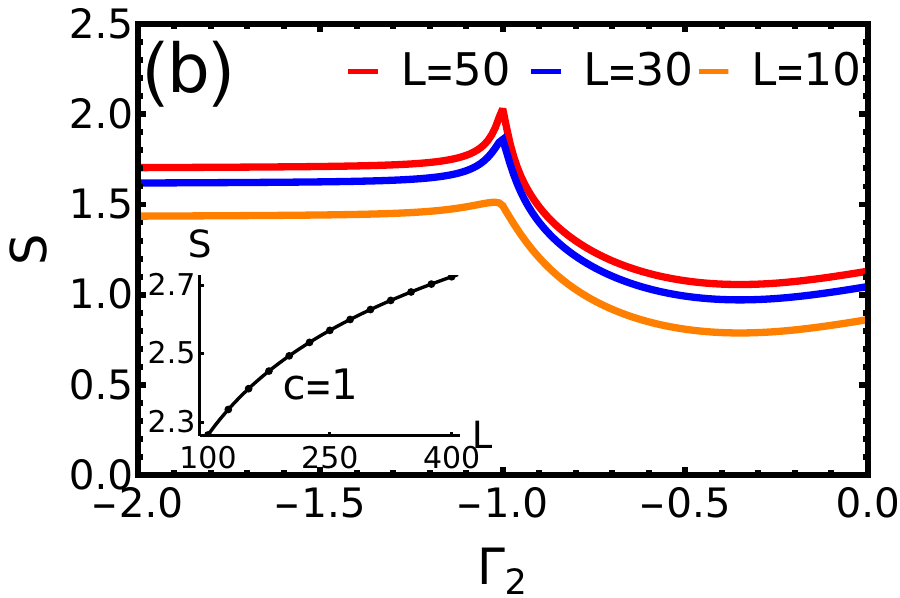}  
	\caption{\label{EE} Entanglement entropy at criticality (with $\Gamma_{0}=1$). Topological transitions are identified with (a) minima at $MC_1$ ($\Gamma_{2}=1$) and (b) maxima at $MC_2$ ($\Gamma_{2}=-1$). Inset shows scaling of $S$ at $MC_{2}$ with central charge $c=1$.}
\end{figure}
\section{Entanglement Entropy}\label{sec6}
The characteristics of a criticality can be effectively quantified from the entanglement entropy (EE) of the ground state by arbitrarily dividing a system into two subsystems~\cite{PhysRevLett.90.227902,PhysRevLett.102.255701,PhysRevLett.121.076802}.  
Taking the advantage of Wick's theorem, the eigenvalues of the
reduced density matrix can be extracted
from the two point correlation matrix, which in the thermodynamic limit can be written as~\cite{PhysRevLett.121.076802}
\begin{equation}
C_{i,j}= \int_{-\pi}^{\pi} \frac{dk}{2\pi} e^{ik(i-j)} \mathcal{G}(k), \;\;\; \text{where} \;\;\;
\mathcal{G}(k)=\frac{\boldsymbol{\chi.\sigma}}{E_k},
\end{equation}
with $1\le i,j \le L$ (where $L$ is the subsystem size).
The EE ($S$) can be computed as~\cite{PhysRevLett.121.076802}
\begin{equation}
S=-\frac{1}{2} \sum_{\kappa=\pm,\Lambda_i} \frac{1-\kappa \Lambda_i}{2} \ln \left( \frac{1-\kappa \Lambda_i}{2}\right) ,
\end{equation}
where $\Lambda_i$ are the eigenvalues of correlation matrix. 
The EE signals the topological transition  at the multicritical points, as shown in Fig.\ref{EE}.
The profile of EE shows, maxima at $MC_2$ (Fig.\ref{EE}(b)) and surprisingly minima at $MC_1$ (Fig.\ref{EE}(a)). 

For the generic model in Eq.\ref{generic-H}, the $MC_1$ is the intersection point of fixed and critical lines. Remarkably, at $MC_1$, we observe that the fixed point characteristic is more dominant which results in the minima of entanglement entropy, in oppose to the critical point behavior where entanglement is supposed to maximize due to the enhanced correlations (see Appendix.\ref{App-EE} for more details).
Besides, at $MC_1$, the bulk is not a CFT. This can be seen from the multiplicity factor ($m$) i.e. the degenerate zeros on the unit circle of the complex function associated with the Hamiltonian. The multiplicity at $MC_1$ is $m=2$ (see Fig.\ref{Winding-zeros}(g,h,i,j)). As shown in Ref.\cite{verresen2018topology}, if the complex function has degenerate zeros with multiplicity $m$, the bulk is not CFT and implies the dynamical exponent $z=m$. This is consistent with $z$ value obtained for $MC_1$ (see Fig.\ref{exponents}(c)).

At $MC_2$, the EE is $S=S_0+(c/3) \log L$~\cite{PhysRevLett.121.076802} where constant $S_0=0.72$ and the central charge $c=1$ as shown in the inset of Fig.\ref{EE}(b). The value of $c$ at $MC_2$ is consistent with Ref.~\cite{PhysRevLett.121.076802}, where $c$ was found to be the sum of the central charges of intersecting criticalities. As $MC_2$ is the intersecting point of the two Ising criticalities ($c=1/2$), we get $c=1$.

\section{Conclusion}\label{sec7}
In this work, we reconstruct various tools to characterize the unusual topological phase transitions between distinct critical phases of an extended model that represents topological insulators and superconductors at criticality.
Bound state solutions of the Dirac equation and winding number defined for criticality show that the transitions between the critical phases occur through multicritical points $MC_{1,2}$ of different universality classes as captured through the critical exponents obtained from the divergence of the curvature function. 
There exists an interesting {\it swapping} behavior of the critical momenta $k_0^{mc}$ at $MC_2$ which manifests in the behavior of curvature function. 
A scaling theory based on the 
curvature function unravels that the transitions at $MC_{1,2}$ can be efficiently identified from the RG flow in the parameter space and also shows that, $MC_{2}$ manifests as unstable fixed line of RG flow for $MC_{1}$ and vice versa. A diverging correlation length obtained from the Wannier state correlation function, which essentially is the Fourier transform of the curvature function, indicates the ocurrences of topological phase transitions at $MC_{1,2}$ . 
Moreover, the unique transitions at $MC_{1,2}$ are characterized with the minima and maxima of entanglement entropy respectively revealing an intriguing dominance of the fixed point over the criticality at $MC_1$.

Our proposed framework, in general, can be applied to the driven systems and 
higher dimensional systems.
A unique advantage of having topological non-trivial criticalities is that 
the quantum information remains robust upon tuning the system towards it~\cite{verresen2019gapless}. By identifying the multicritical points 
one can choose a proper criticality to tune into and avoid the decoherence due to bulk gap closing and opening.
Our topological model at criticality can be simulated with a good control over the tunable parameters in the suitable experimental platforms which include the superconducting circuit with a single qubit~\cite{PhysRevB.101.035109,niu2021simulation} and the ultracold atoms mimicking the topological models ~\cite{goldman2016topological,xie2019topological,an2018engineering,meier2018observation,meier2016observation}, especially the Kitaev model with controlled NN and NNN couplings \cite{kraus2012preparing,jiang2011majorana,an2018engineering}. 

\begin{acknowledgements}
We would like to thank Wei Chen, Griffith M. Rufo, Ruben Verresen, Yuanzhen Chen, Aditi Mitra, Vinod N Rao, Randeep N. C for the useful discussions. RRK, YRK and RS would like to acknowledge DST (Department of Science and Technology, Government of India-EMR/2017/000898 and CRG/2021/00996) and AMEF (Admar Mutt Education Foundation) for the funding and support. NR acknowledges Indian Institute of Science (IISc.), Bangalore for support through the Institution of Eminence (IoE) Post-Doc program. This research was supported in part by the International Centre for Theoretical Sciences (ICTS) during a visit for participating in the program -  Novel phases of quantum matter (Code: ICTS/topmatter2019/12).
\end{acknowledgements}

\appendix
\section{Physical relevance of model Hamiltonian}\label{Model_Supply}
The model considered in Eq.\ref{generic-H} is a generic two band model for spinless fermions in 1D lattice with nearest neighbor (NN) and next nearest neighbor (NNN) coupling amplitudes of electrons. It maps into extended Su-Schrieffer–Heeger (SSH) \cite{PhysRevLett.42.1698,hsu2020topological} and Kitaev chains \cite{kitaev2001unpaired,niu2012majorana} in momentum space, which are the simplest 1D models for topological insulators and superconductors respectively. The tight-binding Hamiltonians can be written as
\begin{align}
H_{SSH} &= \alpha_0 \sum_{i}  c_{i,a}^{\dagger} c_{i,b} + \alpha_1 \sum_{\left\langle ij\right\rangle } (c_{i,a}^{\dagger} c_{j,b} + h.c) \nonumber\\
&+ \alpha_2 \sum_{\left\langle \left\langle ij\right\rangle \right\rangle }( c_{i,a}^{\dagger} c_{j,b} + h.c), \label{SSH}
\end{align}
\begin{align}
H_{Kitaev} &= \beta_0 \sum_{i} ( 2 c_{i}^{\dagger}c_{i}-1) - \beta_1 \sum_{\left\langle ij\right\rangle } (c_{i}^{\dagger}c_{j} + c_{i}^{\dagger}c_{j}^{\dagger} + h.c) \nonumber\\
&- \beta_2 \sum_{\left\langle \left\langle ij\right\rangle \right\rangle } ( c_{i}^{\dagger}c_{j} +  c_{i}^{\dagger} c_{j}^{\dagger} + h.c),
\label{kitaev}
\end{align}
where $c^{\dagger}_{i,j}$ and $c_{i,j}$ are the fermionic creation and annihilation operators. In $H_{SSH}$, the subscripts $a,b$
denote the sub-lattices, with onsite potential $\alpha_0$ and NN (NNN) hopping amplitude $\alpha_{1(2)}$. In $H_{Kitaev}$, $\beta_0$ is onsite potential and $\beta_{1(2)}$ is NN (NNN) pairing and hopping amplitudes. 

\begin{figure}[t]
	\includegraphics[width=4.2cm,height=3.2cm]{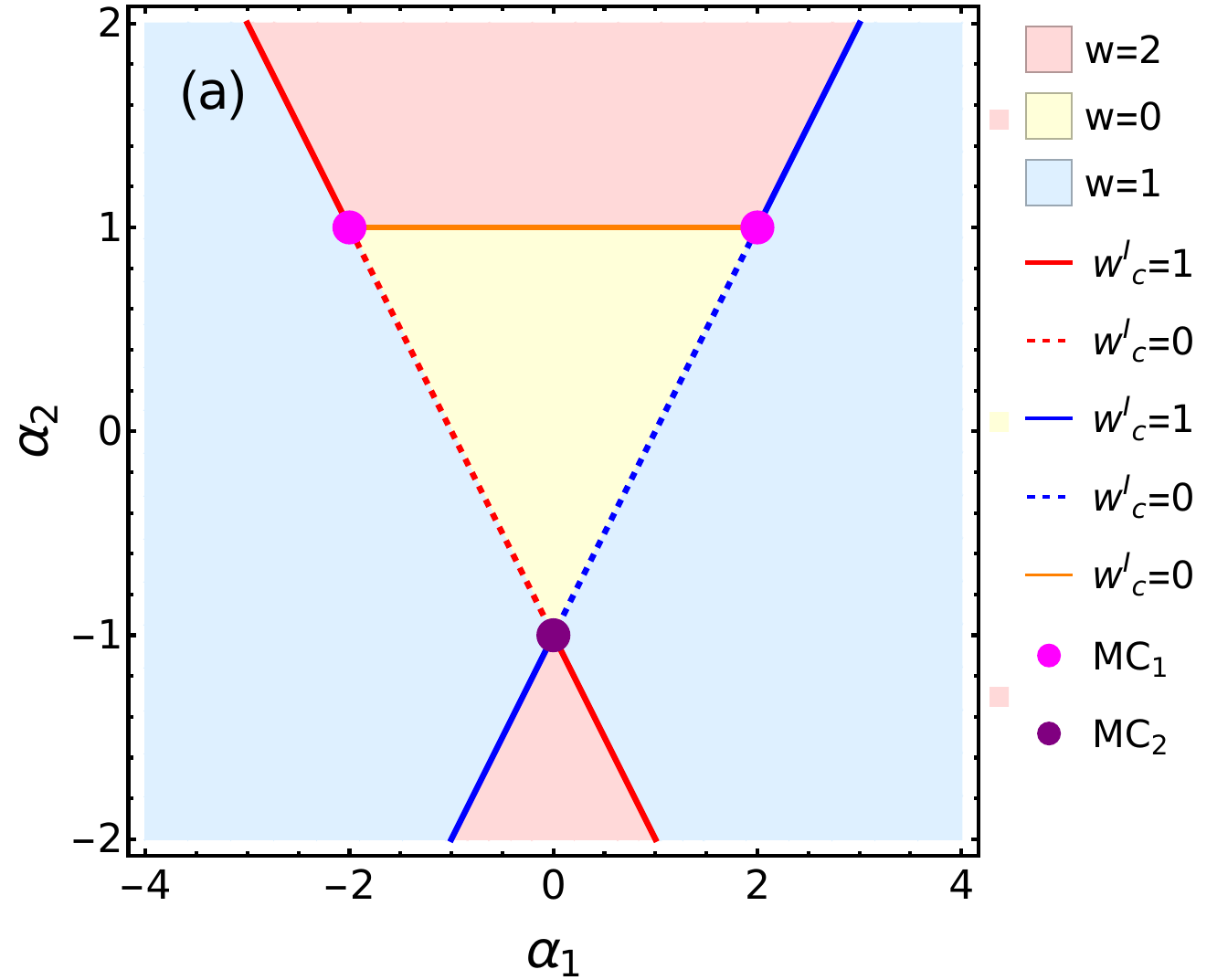}
	\includegraphics[width=4.2cm,height=3.3cm]{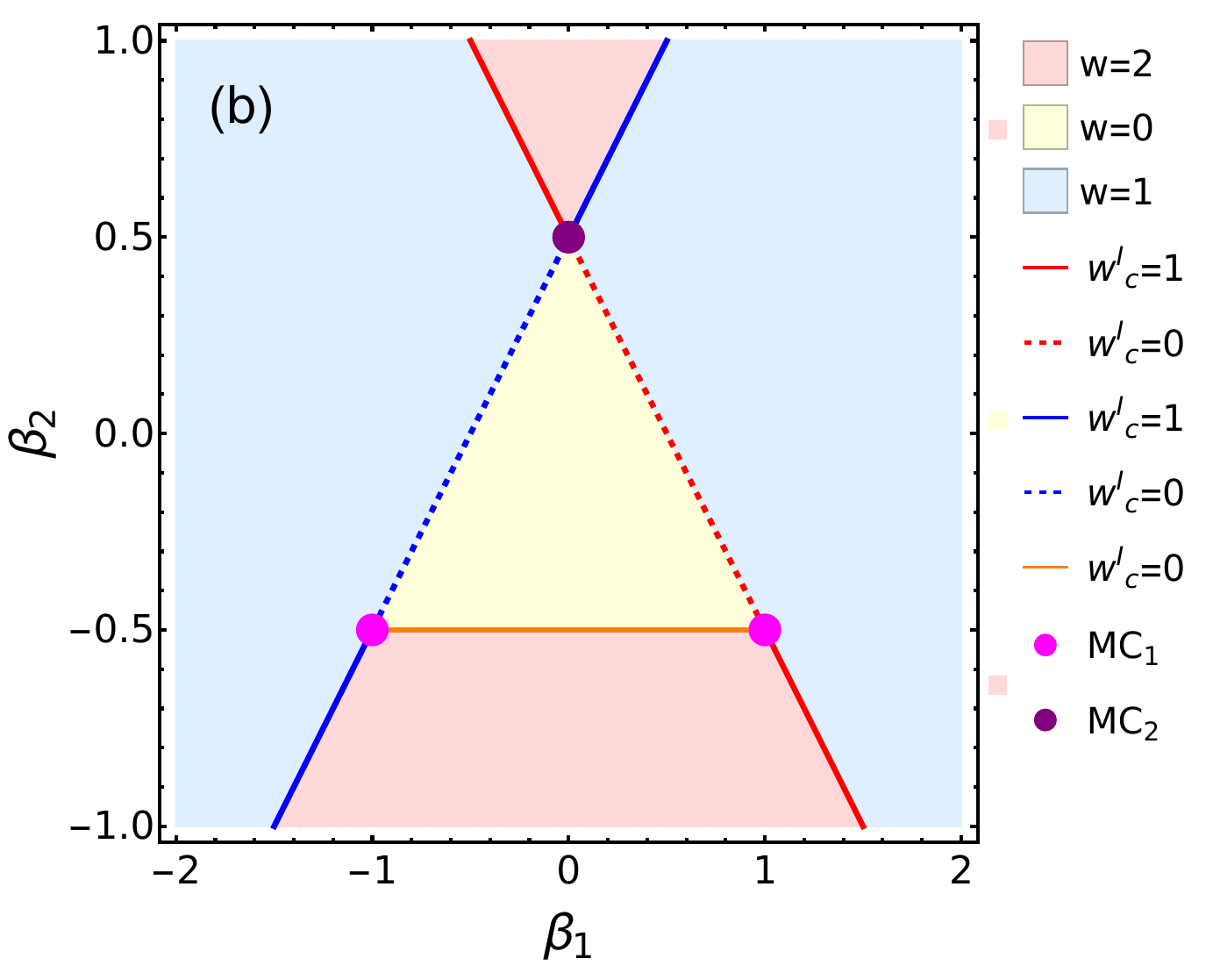}
	\caption{\label{TPD} \small Topological phase diagrams of (a) extended SSH model (for $\alpha_0=1$.) and (b) extended Kitaev model (for $\beta_0=0.5$). High symmetry critical lines for $k_0=0$ and $\pi$ are 
		represented in Red and Blue respectively, while non-high symmetry critical lines are in Orange. The topological trivial and non-trivial critical phases are represented in dashed and solid lines respectively. The $w_c^I$ are the winding numbers at criticality (Eq.\ref{wind-integer}). The critical phases are separated by multicritical points
		$MC_1$ (magenta dots) and $MC_2$ (purple dots). Each high symmetry criticalities allows topological transition between distinct critical phases through multicritical points.
	}
\end{figure}

The Hamiltonians can be readily diagonalised by Fourier transformation 
to obtain a generalized Bloch Hamiltonian in the basis of spinor $\psi_k$ 
\begin{equation}
H_{SSH} = \sum_{k} \psi^{\dagger}_k \mathcal{H}_{SSH} \psi_k \quad \text{with} \quad \psi_k=\begin{pmatrix}
c_{a,k} & c_{b,k} 
\end{pmatrix}^T
\end{equation}
The Hamiltonian $\mathcal{H}_{SSH}(k) = \chi_{x}.\sigma_x+\chi_{y}.\sigma_y$, where $\chi_{x}= \alpha_0 + \alpha_1 \cos k + \alpha_2 \cos 2k$ and $\chi_{y}= \alpha_1 \sin k + \alpha_2 \sin 2k$.
The excitation spectra can be obtained as 
$E_k=\pm \sqrt{\chi_{x}^2+\chi_{y}^2}$. The gap closing points (i.e., $E_k=0$) for a specific $k_0$ defines critical surfaces or 
phase boundaries which separate topologically distinct gapped phases. The gapless edge excitations of these gapped phases are quantified in terms of
winding number $w$, which counts the number of edge modes present in the corresponding gapped phases.
There are three critical surfaces for extended SSH model. Two of them are with high symmetry nature (i.e, $k_0=-k_0$), $\alpha_1=-(\alpha_0+\alpha_2)$ and $\alpha_1=(\alpha_0+\alpha_2)$ respectively for $k_0=0$ and $\pi$. One with non-high symmetry nature (i.e, $k_0\neq-k_0$), $\alpha_0=\alpha_2$ for $k_0=\cos^{-1}(-\alpha_1/2\alpha_2)$. Without loss of generality, we assume $\alpha_0=1$, hence critical surfaces and multicritical lines will be critical lines and multicritical points respectively on the $\alpha_1-\alpha_2$ plane, as shown in Fig.\ref{TPD}(a). The three critical lines distinguish the gapped phases with invariant number $w=0,1,2$. There are three multicritical points named $MC_1$ (two of them) and $MC_2$, with distinct nature, at which the critical lines meet~\cite{malard2020multicriticality}. 

The edge mode remains localized at the criticalities (critical lines) between the topological non-trivial gapped phases ($w=1$ and $w=2$), which give rise to the topological characteristics to the criticality. The same does not occur at the criticality between a trivial and non-trivial gapped phases ($w=0$ and $w=1$). This results in the criticality to get separated into two distinct critical phases with trivial and non-trivial topological properties. 
The multicritical points $MC_{1,2}$,
with quadratic (i.e. $E_k \propto k^2$) and linear dispersions (i.e. $E_k \propto k$) respectively, facilitates the topological transition at criticality between trivial and non-trivial critical phases.

Similar qualitative behavior can also be observed in the extended Kitaev model due to the striking similarity in the phase diagram with SSH model. For the Kitaev model one can obtain
\begin{equation}
H_{Kitaev} = \sum_{k} \psi^{\dagger}_k \mathcal{H}_{Kitaev} \psi_k \quad \text{with} \quad \psi_k=\begin{pmatrix}
c_{k} & c^{\dagger}_{-k} 
\end{pmatrix}^T
\end{equation}
The Hamiltonian $\mathcal{H}_{Kitaev}(k) = \chi_{x}.\sigma_x+\chi_{y}.\sigma_y$, where $\chi_{x}= 2\beta_0 - 2\beta_1 \cos k - 2\beta_2 \cos 2k$ and $\chi_{y}= 2\beta_1 \sin k + 2\beta_2 \sin 2k$, 
after a rotation along $\sigma_y$.
The gap closing critical surfaces for this case are $\beta_1=-(\beta_0-\beta_2)$, $\beta_1=(\beta_0-\beta_2)$ and $\beta_0=-\beta_2$ respectively for $k_0=0$, $k_0=\pi$ and $k_0=\cos^{-1}(-\beta_1/2\beta_2)$. These phase boundaries 
separate the gapped phases with invariant numbers $w=0,1,2$ as shown in Fig.\ref{TPD}(b) (for $\beta_0=0.5$). Localized edge modes living at the criticalities between the non-trivial topological gapped phases can be observed here as well which defines trivial and non-trivial critical phases with distinct topological properties. The multicritical points $MC_{1,2}$ mediate the topological transition at criticality between critical phases with distinct topological nature and
share the same properties as in the case of SSH model.

To study the unusual topological transition at criticalities we consider a generic model which essentially summarize both
SSH and Kitaev model, thereby giving one platform to study both topological insulator and superconductor models in one dimension.
We define a generalized Bloch Hamiltonian for two band model by setting $\alpha_0=2\beta_0=\Gamma_{0}$, $\alpha_1=-2\beta_1=\Gamma_{1}$, and $\alpha_2=-2\beta_2=\Gamma_{2}$. This model captures the physics of both SSH and Kitaev models, especially the phenomenon of multicriticality and the corresponding topological transition.

\section{Numerical results of edge modes and topological transition at criticality}\label{Numerical}
\begin{figure}[t]
	\includegraphics[width=3.7cm,height=3.2cm]{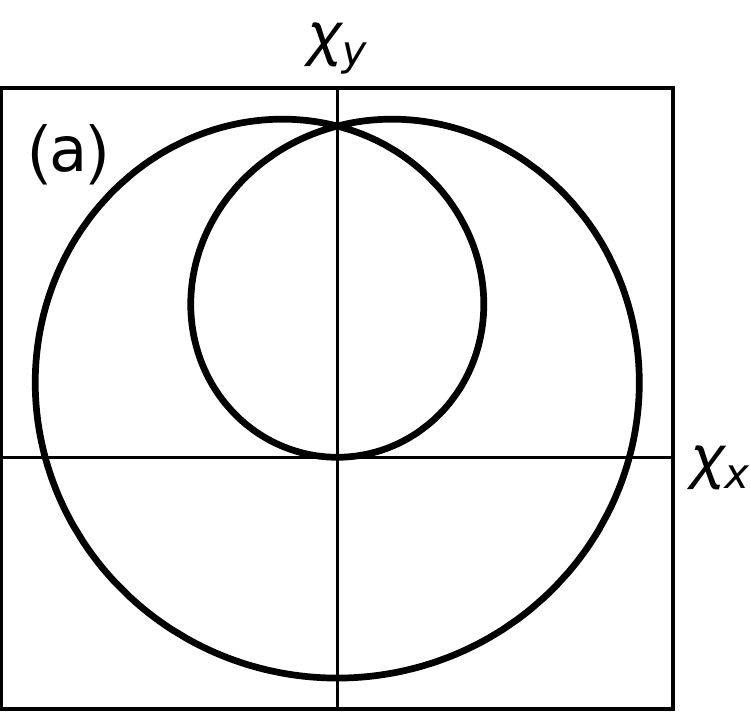}
	\hspace{0.2cm}
	\includegraphics[width=3.7cm,height=3.2cm]{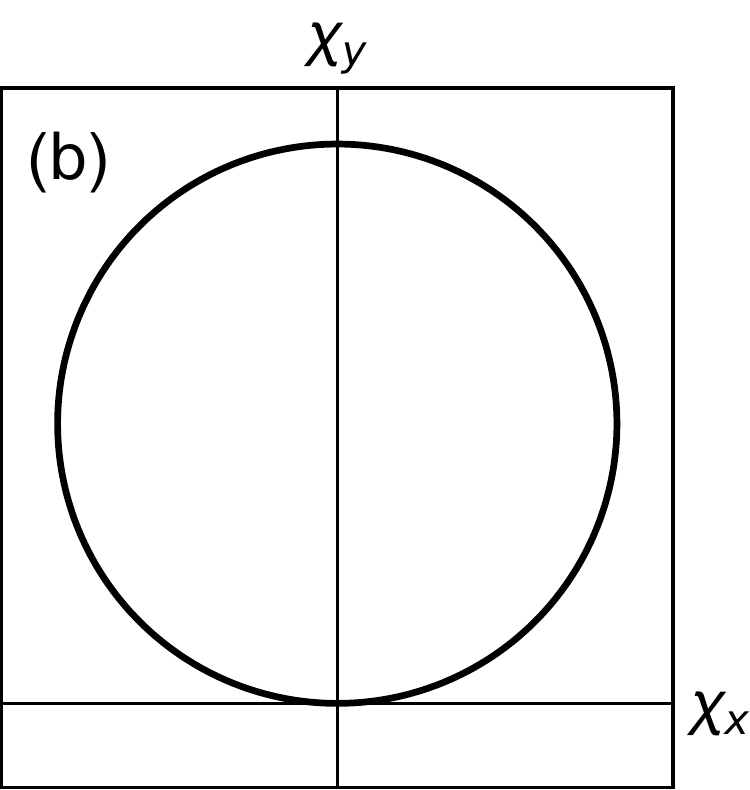} 
	\caption{\label{SV}\small Pseudo spin-vector at criticality. (a) Non-trivial critical phase. (b) Trivial critical phase.}
\end{figure}

We begin by discussing the behavior of the pseudo spin-vectors to identify the trivial and non-trivial criticalities. The characteristic feature of the parameter space curve at criticality is that it
passes through the origin while tracing closed curve. Non-trivial critical phases can be identified with the
emergence of secondary loops 
which
encircle the origin indicating a finite winding number or edge modes at criticality, as shown in Fig.\ref{SV}(a). In trivial critical phase parameter space curves always passes through the origin without encircling loops, thus there is no edge modes at criticality, as shown in Fig.\ref{SV}(b).
\begin{figure}[t]
	\begin{minipage}[t]{0.45\columnwidth}
		\includegraphics[width=4.2cm,height=3cm]{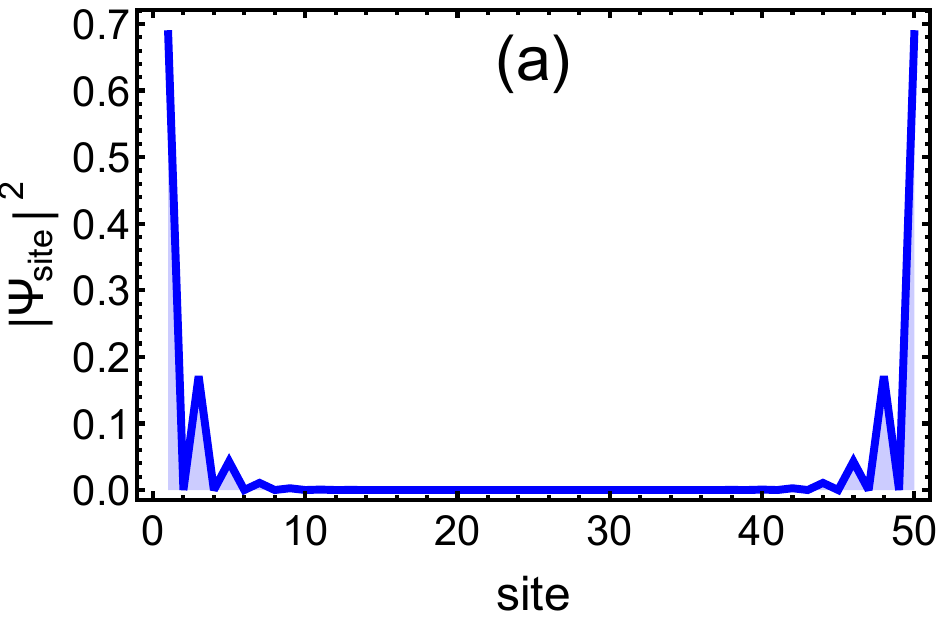} \\
		\vspace{0.3cm}
		\includegraphics[width=4.2cm,height=3cm]{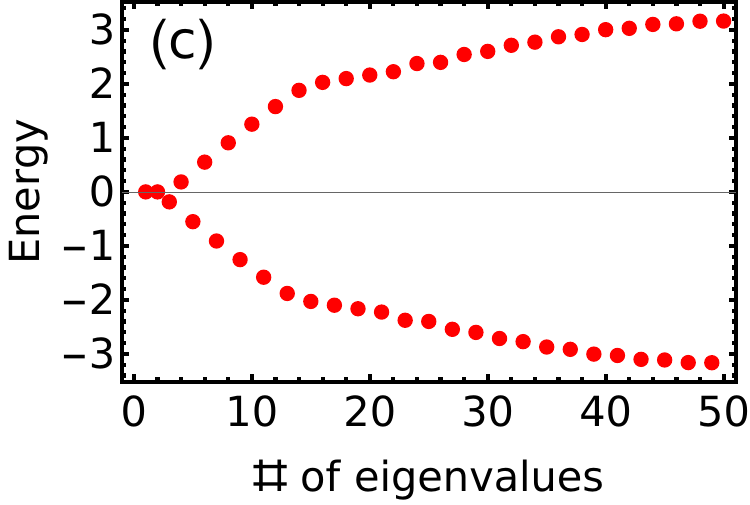} \\
		\vspace{0.3cm}
		\includegraphics[width=4.2cm,height=3cm]{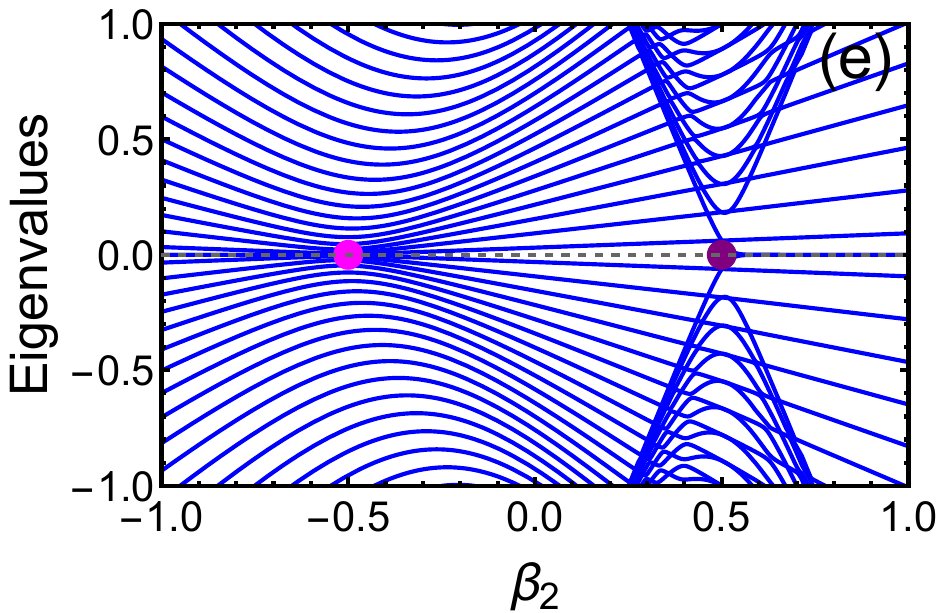} 
	\end{minipage}
	\hfil
	\begin{minipage}[t]{0.45\columnwidth}
		\includegraphics[width=4.2cm,height=3cm]{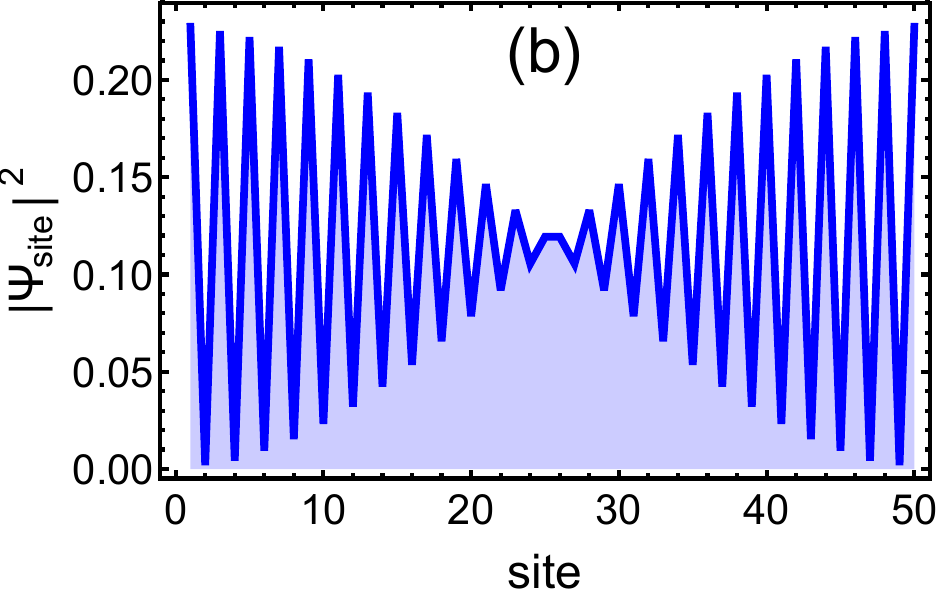} \\
		\vspace{0.3cm}
		\includegraphics[width=4.2cm,height=3cm]{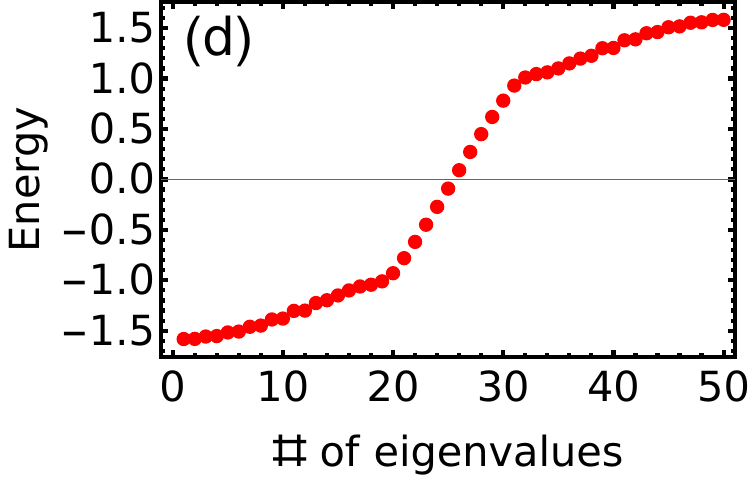} \\
		\vspace{0.3cm}
		\includegraphics[width=4.2cm,height=3cm]{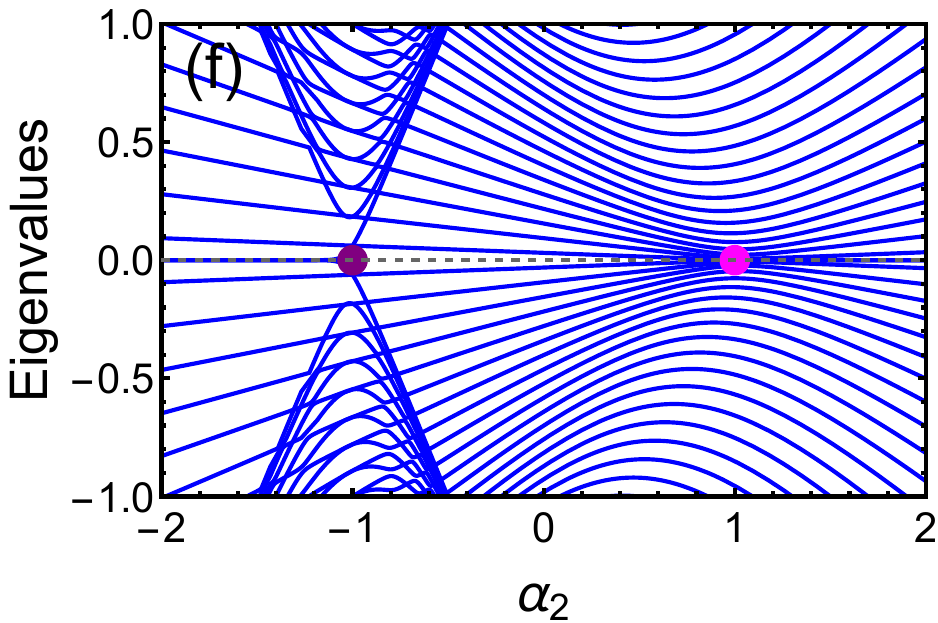} 
	\end{minipage}
	\caption{\label{Topo-Trans}\small Numerical results for edge mode and topological phase transition at criticality. Probability distribution at (a) non-trivial critical phase and (b) trivial critical phase. Eigenvalue distribution at (c) non-trivial critical phase and (d) trivial critical phase. Energy spectra at criticalities with respect to the parameters: (e) $\beta_{2}$ (for critical Kitaev model) and (f) $\alpha_2$ (for critical SSH model). The milticritical points are represented as magenta and purple dots. Zero energy states are, present at non-trivial critical phases and absent at trivial critical phase. The multicritical points differentiate the trivial and non-trivial phases.}
\end{figure}

Numerical diagonalization of the Hamiltonians in Eq.\ref{SSH} and Eq.\ref{kitaev} (the results shown in this section summerizes both SSH and Kitaev model in open boundary condition) reveals that for the non-trivial critical phases the probability of wave function
significantly distributes at the edge of the finite open chain representing the stable localized edge modes, as shown in Fig.\ref{Topo-Trans}(a). The corresponding eigenvalue distribution shows two of the eigenvalues trapped at zero energy even if there is no bulk gap, as shown in Fig.\ref{Topo-Trans}(c).
In case of the trivial critical phase, probability distribution can be found delocalized over the entire system, as shown in Fig.\ref{Topo-Trans}(b). Correspondingly, there are no eigenvalues living at zero energy, as shown in Fig.\ref{Topo-Trans}(d). The localization and delocalization of the edge modes change across the multicritical points $MC_{1,2}$ which thus differentiate between trivial and non-trivial critical phases. 

The topological transition among the non-trivial and trivial critical phases can be identified in the energy spectrum with the system parameter. The presence (absence) of zero energy states dictates the non-triviality (triviality) with respect to the system parameter, as shown in Fig.\ref{Topo-Trans}(e) and (f). Note that, there is no bulk gap in the spectrum since the system is at criticality. The zero energy states represent localized stable edge modes living at the critical phases. The transition among the trivial and non-trivial phases can be seen at the multicritical points.

\section{Bound state solution of Dirac equation for gapped phases}\label{dirac-gapped}
The model Hamiltonian in Eq.\ref{generic-H}
can be recasted in the form of Dirac Hamiltonian in 1D, which represents the topological insulator and superconductor models.
The Dirac Hamiltonian of the model can be obtained by the 
second order expansion of $\boldsymbol{\chi}$ around the gap closing momenta $k_0$
\begin{equation}
\mathcal{H}(k) \approx (m - \epsilon_1 k^2)\sigma_x +  \epsilon_2 k \sigma_y.
\end{equation}
For $k_0=0$ we have $m=(\Gamma_{0}+\Gamma_{1}+\Gamma_{2})$, $\epsilon_1=(\Gamma_{1}+4\Gamma_{2})/2$ and $\epsilon_2=(\Gamma_{1}+2\Gamma_{2})$. For 
$k_0=\pi$, $m=(\Gamma_{0}-\Gamma_{1}+\Gamma_{2})$, $\epsilon_1=(4\Gamma_{2}-\Gamma_{1})/2$ and $\epsilon_2=(2\Gamma_{2}-\Gamma_{1})$.
The continuum version of the model reads (with $\hbar=1$)
\begin{equation}
\mathcal{H}(-i\partial_x) \approx (m + \epsilon_1\partial_x^2)\sigma_x +  (-i \epsilon_2 \partial_x)\sigma_y.
\end{equation}
\begin{figure}[t]
	\includegraphics[width=4.2cm,height=2.7cm]{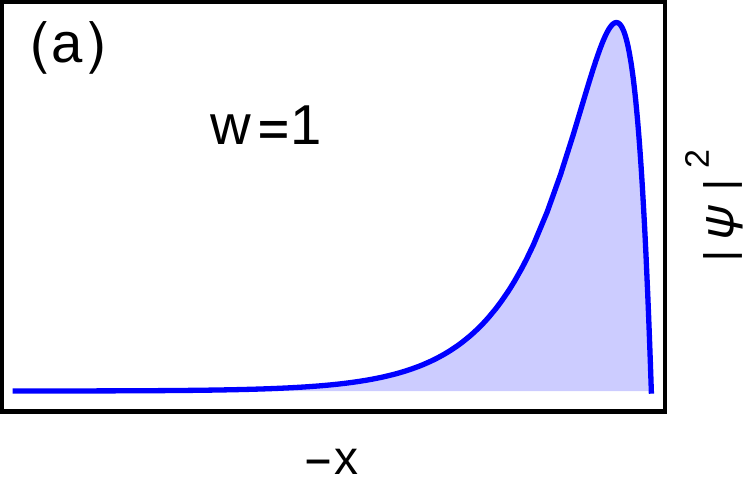}  
	\includegraphics[width=4.2cm,height=2.7cm]{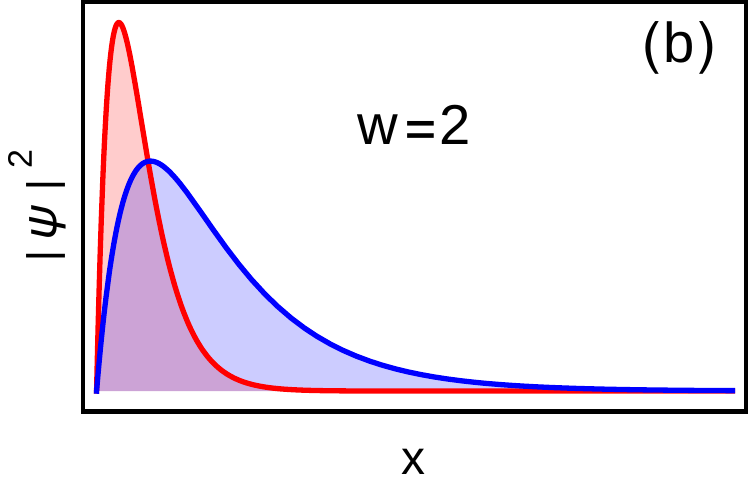}  
	\caption{\label{Dirac-gapped}\small Bound state solution of Dirac equation for gapped phases. (a) Represents the gapped phase with one edge mode $w=1$. (b) Represents the gapped phase with two edge modes $w=2$. Edge modes solutions are localized at the boundary with the localization length $\xi_{\pm}$.}
\end{figure} 
To obtain zero energy solution $\mathcal{H}\psi(x)=0$, we multiply $\sigma_y$ from right-hand side. This implies
the wavefunction $\psi(x)=\rho_{\eta}\phi(x)$, is an eigenstate of $\sigma_z\rho_{\eta}=\eta\rho_{\eta}$. The resulting 
second order
differential equation can be written as
\begin{equation}
\partial^2_{x}\phi(x)= \frac{-(\epsilon_2\partial_{x}+\eta m)\phi(x)}{\eta\epsilon_1}
\end{equation} 
We set the trial wavefunction $\phi(x) \propto e^{x/\xi}$ to obtain the secular equation, which yields the inverse of the decay length 
\begin{equation}
\xi^{-1}_+ \approx - \frac{m}{|\epsilon_2|}.
\end{equation}
The decay length is positive if $m<0$ which identifies the gapped topological non-trivial phase with $w=1$. Similarly,
topological phase with $w=2$ can also be identified by using the ansatz $\phi(x) \propto e^{-x/\xi}$, which under the condition
$m\epsilon_1>0$ yields the decay length $\xi_-\approx |\epsilon_2|/m$. The decay length is positive if $m>0$. 
Even though, the topological trivial phase with $w=0$ is also identified with $m>0$, it does not host any zero energy
solution since the region $m>0$ for trivial phase satisfies the relation $m\epsilon_1<0$. If the parameter $m\epsilon_1<0$,
spin distribution of the ground state does not show anti-parallel spin orientation in momentum space \cite{shun2018topological}. If  $m\epsilon_1>0$
is satisfied, spin orientation align in the opposite directions with the increasing momentum. Thus the gapped phases $w=2$ and
$w=0$ are identified with the condition $m\epsilon_1\lessgtr 0$ respectively. 
The wave-function for zero energy solution can be derived to be
\begin{equation}
\psi(x)  \propto \psi(0) (e^{\pm x/\xi_+}- e^{\pm x/\xi_-}),
\end{equation}
up to normalization constant. The solution is exponentially localized near the boundary, as shown in Fig.\ref{Dirac-gapped} for
different gapped phases.

\section{Analytical evaluation of critical exponents}\label{expo-nume}
The critical exponents in Section.\ref{expo} can also be estimated analytically by writing the curvature function (Eq.\ref{curv-crit12}) in Ornstein-Zernike form. It can be achieved by expanding the  
pseudo-spin vector $\boldsymbol{\chi}(k)$ around $k_0^{mc}$ up to third order.
\begin{align}
\boldsymbol{\chi}(k)|_{k=k_0^{mc}} &\approx \boldsymbol{\chi}(k_0^{mc}) + \partial_k\boldsymbol{\chi}(k_0^{mc}) \delta k +\frac{\partial_k^2\boldsymbol{\chi}(k_0^{mc})}{2} \delta k^2 \nonumber\\
&+ \frac{\partial_k^3\boldsymbol{\chi}(k_0^{mc})}{6} \delta k^3,
\end{align}
Expansion of the individual components of the vectors $\chi_x(k)|_{k=k_0^{mc}}= \Gamma_0 (1 \pm \cos k) + \Gamma_2 (\cos 2k \pm \cos k)$ and $\chi_y(k)|_{k=k_0^{mc}} = \Gamma_2 (\sin 2k \pm \sin k)\pm \Gamma_0 \sin k $ for both the criticalities of the 
model 
yields 
\begin{align}
\text{For} \;\;\; MC_1: \;\;\;
\chi_x(k)|_{k=k_0^{mc}}&\approx \frac{(\Gamma_0 - 3\Gamma_2)}{2} \delta k^2.\\
\chi_y(k)|_{k=k_0^{mc}}&\approx (\Gamma_2-\Gamma_0) \delta k + \frac{\Gamma_0 - 7\Gamma_2}{6} \delta k^3.\\
\text{For} \;\;\; MC_2: \;\;\;
\chi_x(k)|_{k=k_0^{mc}}&\approx (\Gamma_0 + 3\Gamma_2) \delta k.\\
\chi_y(k)|_{k=k_0^{mc}}&\approx 2(\Gamma_2+\Gamma_0) + \frac{\Gamma_0 + 5\Gamma_2}{2} \delta k^2.
\end{align}
The expression for $MC_2$ is obtained after considering the swapping of $k_0^{mc}$. The Ornstein-Zernike form of the curvature function for $MC_1$ can be obtained as
\begin{align}
F(k,\delta \mathbf{\Gamma}_{c})&= \frac{\chi_y\partial_k \chi_x - \chi_x \partial_k \chi_y}{\chi_x^2+\chi_y^2} \nonumber\\
&= \frac{\left( \frac{A \delta \mathbf{\Gamma}_{c} \delta k^2 - A B \delta k^4}{\delta \mathbf{\Gamma}^2_{c} \delta k^2} \right) }{1+\left( \frac{A^2+2 \delta \mathbf{\Gamma}_{c} B}{\delta \mathbf{\Gamma}^2_{c}} \right)\delta k^2 + \left( \frac{B^2}{\delta \mathbf{\Gamma}^2_{c}}\right)  \delta k^4 } \nonumber\\
&=\frac{F(k_0^{mc},\delta\mathbf{\Gamma}_{c})}{1+\xi_{c}^2\delta k^2+\xi_{c}^4\delta k^4},
\end{align}
where $\delta \mathbf{\Gamma}_{c}=|\mathbf{\Gamma}_{c}-\mathbf{\Gamma}_{mc}|=( \Gamma_2-\Gamma_0)$, $A= (\Gamma_0- 3\Gamma_2)/2$ and $B= (\Gamma_0- 7\Gamma_2)/6$.
Similarly, for $MC_2$ it reads	
\begin{align}
F(k,\delta \mathbf{\Gamma}_{c})&= \frac{\chi_y\partial_k \chi_x - \chi_x \partial_k \chi_y}{\chi_x^2+\chi_y^2} \nonumber\\
&= \frac{\left( \frac{A \delta \mathbf{\Gamma}_{c} - A B \delta k^2}{\delta \mathbf{\Gamma}^2_{c}} \right) }{1+\left( \frac{A^2+2 \delta \mathbf{\Gamma}_{c} B}{\delta \mathbf{\Gamma}^2_{c}} \right)\delta k^2 + \left( \frac{B^2}{\delta \mathbf{\Gamma}^2_{c}}\right)  \delta k^4 } \nonumber\\
&= \frac{F(k_0^{mc},\delta\mathbf{\Gamma}_{c})}{1+\xi_{c}^2\delta k^2+\xi_{c}^4\delta k^4},
\end{align}
where $\delta \mathbf{\Gamma}_{c}=2( \Gamma_2+\Gamma_0)$, $A= (\Gamma_0+ 3\Gamma_2)$ and $B= (\Gamma_0+ 5\Gamma_2)/2$. Now the critical exponents can be obtained using Eq.\ref{critical-exponents-crit}. The exponent $\gamma$ is given by
\begin{equation}
F(k_0^{mc},\delta\mathbf{\Gamma}_{c}) = A \delta \mathbf{\Gamma}^{-1}_{c} \implies \gamma =1.
\end{equation}
Exponent $\nu$ can be obtained by identifying the dominant term among the coefficients of $\delta k^2$ and $\delta k^4$. It can be easily seen that approaching multicritical points $MC_{1,2}$ on both the criticalities yields $A>\sqrt{2 B},\sqrt{B}$. This implies
\begin{equation}
\xi_{c} = A \delta \mathbf{\Gamma}^{-1}_{c} \implies \nu =1.
\end{equation}
Thus both the numerical and analytical methods yield the same values of critical exponents for topological transition through multicritical points at criticality.
\begin{figure}[t]
	\includegraphics[width=5cm,height=4.7cm]{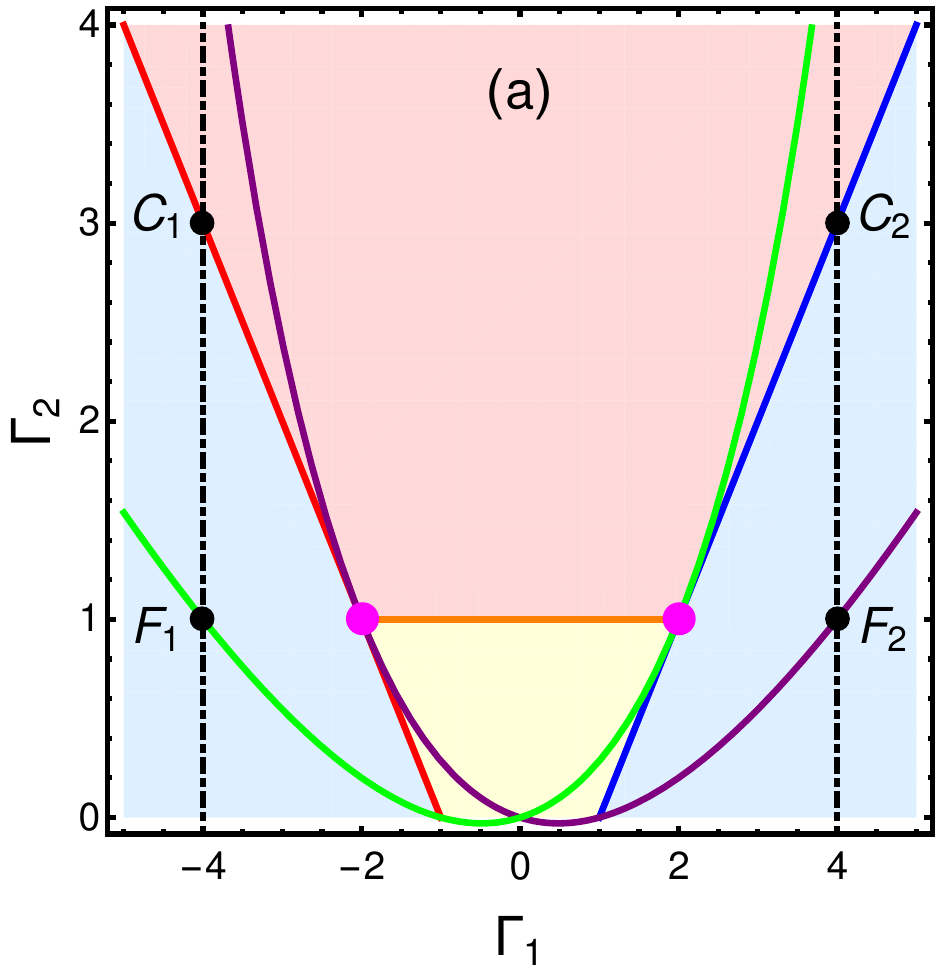} 
	\includegraphics[width=7cm,height=5cm]{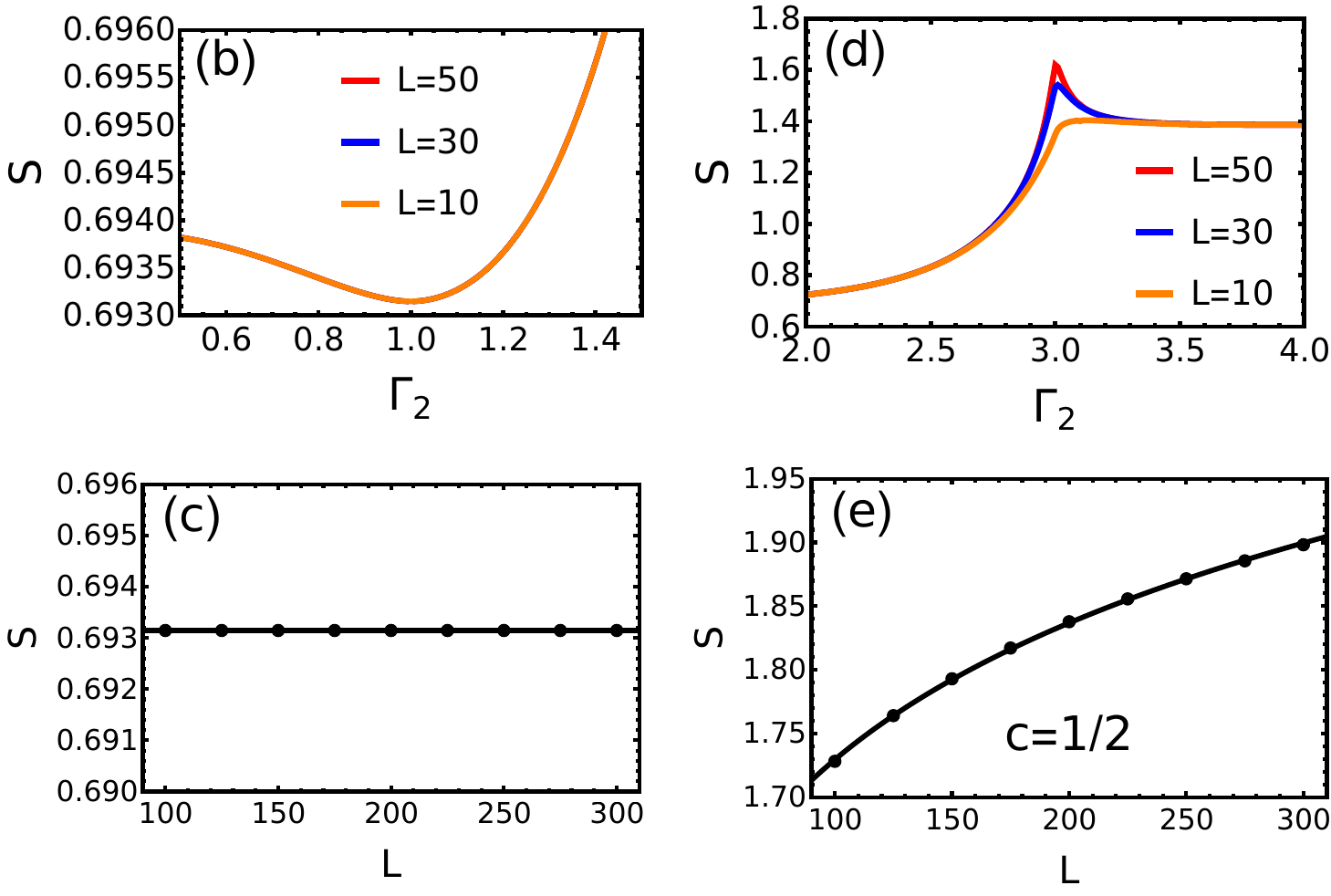}  
	\caption{\label{EES} EE at critical and fixed points. (a) Phase diagram of the model in Eq.\ref{generic-H} (same as in Fig.\ref{TPD}(a)). It is plotted in $\Gamma_{1}-\Gamma_{2}$ plane with $\Gamma_{0}=1$. The critical lines are represented as Blue, Red and Orange lines. The multicritical points $MC_1$ are the magenta dots at $(\Gamma_1,\Gamma_{2})=(\pm 2, 1)$. These points are the intersection points of the fixed lines (represented in green and purple) and high symmetry critical lines (red and blue lines). To see the EE profile at fixed and critical points, we choose two paths at $\Gamma_{1}=\pm 4$ (black dashed lines). The points $C_1,C_2$ and $F_1,F_2$ are the critical and fixed points respectively, along the two paths. The behavior of critical (fixed) points $C_1$ ($F_1$) and $C_2$ ($F_2$) are identical, therefore only the plots for $C_1$ and $F_1$ are shown. (b) Variation EE with parameter in the vicinity of $F_1$ $(\Gamma_1=-4,\Gamma_{2}=1)$ for different subsystem sizes. (c) Scaling of EE at the fixed point $F_1$ the subsystem size $L$. (d) EE in the vicinity of the point $C_1$  i.e. $(\Gamma_1=-4,\Gamma_{2}=3)$.  (e) Scaling of EE at the critical point $C_1$. The EE increases with the subsystem size as $S=S_0+(c/3)\log(L)$ where the constant $S_0=1.01$ and central charge $c=1/2$ representing Ising criticalities.}
\end{figure}
\section{Entanglement entropy at critical and fixed points}\label{App-EE}
In order to understand the behavior of EE at the multicritical point $MC_1$, 
at first, we show $MC_1$ is the intersection point of fixed and critical lines in the parameter space. The fixed lines can be obtained by performing the curvature function renormalization group method~\cite{chen2016scaling}, to capture the topological transition between gapped phases of the generic model in Eq.\ref{generic-H}, in $\Gamma_{1}-\Gamma_{2}$ plane with $\Gamma_{0}=1$.
There are two high symmetry critical lines $\Gamma_{2}=(-\Gamma_{0}\mp \Gamma_{1})$ for $k_0=0,\pi$ respectively. The CRG equations can be obtained as
\begin{align}
\frac{d\Gamma_{1}}{dl}&=-\frac{\Gamma_{1}(\Gamma_{2}\mp\Gamma_{1})+\Gamma_{0}(\Gamma_{1}\pm 8\Gamma_{2})}{\Gamma_{0}\pm\Gamma_{1}+\Gamma_{2}},\\
\frac{d\Gamma_{2}}{dl}&=\mp\frac{(\Gamma_{0}-\Gamma_{2})
	(\Gamma_{1}(\Gamma_{2}\mp\Gamma_{1})+\Gamma_{0}(\Gamma_{1}\pm 8\Gamma_{2}))}{(2\Gamma_{0}\pm\Gamma_{1})(\Gamma_{0}\pm\Gamma_{1}+\Gamma_{2})},
\end{align}
where the upper and lower signs are for $k_0=0,\pi$ respectively. The critical and fixed lines can be obtained from the CRG equations using the conditions $|d\mathbf{\Gamma}/dl|\rightarrow\infty$ and $|d\mathbf{\Gamma}/dl|\rightarrow0$ respectively~\cite{chen2018weakly}, and are depicted in Fig.\ref{EES}(a). From the CRG equations for $k_0=0$, the fixed line can be obatined as $\Gamma_{2}=(\Gamma_{1}^2- \Gamma_{0}\Gamma_{1})/(8\Gamma_{0}+\Gamma_{1})$ (purple line), which intersects the critical line $\Gamma_{2}=-(\Gamma_{0}+\Gamma_{1})$ (red line) at the multicritical point $(\Gamma_{1},\Gamma_{2})=(-2,1)$. Similarly, from the CRG equations for $k_0=\pi$, the fixed line $\Gamma_{2}=(\Gamma_{1}^2+ \Gamma_{0}\Gamma_{1})/(8\Gamma_{0}-\Gamma_{1})$ (green line) is obtained and it intersects the critical line $\Gamma_{2}=(-\Gamma_{0}+\Gamma_{1})$ (blue line) at the multicritical point $(\Gamma_{1},\Gamma_{2})=(2,1)$. Both the multicritical points $(\Gamma_{1},\Gamma_{2})=(\pm2,1)$ are of type $MC_1$ as explained in Fig.\ref{TPD}. Therefore, it is clear that the $MC_1$ is the intersection point of fixed and critical lines.

The EE shows minima at a fixed point in contrast to a critical point (where the EE is maximum), as shown in Fig.\ref{EES}(b,c,d,e). We choose two vertical paths in the parameter space at $\Gamma_{1}=\pm 4$, as shown in Fig.\ref{EES}(a).  
The paths intersect the critical points at $C_1,C_2$ (i.e. at $\Gamma_{2}=3$)
and fixed points at $F_1,F_2$ (i.e. at $\Gamma_{2}=1$).
The EE shows minima as a consequence of minimal correlations at the fixed points, as shown in Fig.\ref{EES}(b) and subsytem-size independence at the fixed points~\ref{EES}(c). Similarly, the maximum correlations
yields the maxima 
at the critical points, as shown in Fig.\ref{EES}(d). 
Moreover, at the critical points the EE is $S=S_0+(c/3) \log L$ where `$c$' is the central charge of the CFT and $S_0$ is a constant. Fig.\ref{EES}(e) shows the scaling of EE at $C_1$ and $C_2$ with $c=1/2$, representing Ising criticality~\cite{verresen2019gapless}. In contrast, the fixed points $F_1$ and $F_2$ are in the gapped phase and the EE remains constant with subsystem size $L$ representing the area law, as shown in Fig.\ref{EES}(c). This demonstrates the distinction between fixed and critical points in terms of EE and its scaling.


\begin{thebibliography}{10}
	\bibitem{haldane1988model}
	F. D. M. Haldane.
	\href{https://journals.aps.org/prl/abstract/10.1103/PhysRevLett.61.2015}{\newblock {\em Phys. Rev. Lett.} \textbf{61}, 2015 (1988).}
	
	\bibitem{kane2005quantum}
	 Charles L Kane and Eugene J Mele.
	\href{https://journals.aps.org/prl/abstract/10.1103/PhysRevLett.95.226801}{\newblock {\em Phys. Rev. Lett.} \textbf{95}, 226801 (2005).}
	
	\bibitem{narang2021topology}
	Prineha Narang, Christina A. C. Garcia and Claudia Felser.
	\href{https://www.nature.com/articles/s41563-020-00820-4}{\newblock {\em Nat. Mater.} \textbf{20}, 293--300 (2021).}
	
	\bibitem{wang2017topological}
	Jing Wang and Shou-Cheng Zhang.
	\href{https://www.nature.com/articles/nmat5012}{\newblock {\em Nat. Mater.} \textbf{16}, 1062--1067 (2017).}
	
	\bibitem{thouless1982quantized}
	D. J. Thouless, M. Kohmoto, M. P. Nightingale, and M. den Nijs.
	\href{https://journals.aps.org/prl/abstract/10.1103/PhysRevLett.49.405}{\newblock {\em Phys. Rev. Lett.} \textbf{49}, 405 (1982).}
	
	\bibitem{hasan2010colloquium}
	M~Zahid Hasan and Charles~L Kane.
	\href{https://journals.aps.org/rmp/abstract/10.1103/RevModPhys.82.3045}{\newblock {\em Rev. Mod. Phys.} \textbf{82}, 3045 (2010).}
	
	\bibitem{continentino2020finite}
	Mucio~A Continentino, Sabrina Rufo, and Griffith~M Rufo.
	\newblock Finite size effects in topological quantum phase transitions.
	\href{https://link.springer.com/chapter/10.1007/978-3-030-35473-2_12}{\newblock {\em Strongly Coupled Field Theories for Condensed Matter and Quantum Information Theory}}, Springer Proceedings in Physics 239 (2020).
	
	\bibitem{verresen2018topology}
	Ruben Verresen, Nick~G Jones, and Frank Pollmann.
	\href{https://journals.aps.org/prl/abstract/10.1103/PhysRevLett.120.057001}{\newblock {\em Phys. Rev. Lett.} \textbf{120}, 057001 (2018).}
	
	\bibitem{verresen2019gapless}
	Ruben Verresen, Ryan Thorngren, Nick~G Jones, and Frank Pollmann.
	 \href{https://journals.aps.org/prx/abstract/10.1103/PhysRevX.11.041059}{\newblock {\em Phys. Rev. X.} \textbf{11}, 041059 (2021).} 
	
	\bibitem{jones2019asymptotic}
	Nick~G Jones and Ruben Verresen.
	\href{https://link.springer.com/article/10.1007/s10955-019-02257-9}{\newblock {\em J. Stat. Phys.} \textbf{175}, 1164--1213 (2019).}
	
	\bibitem{verresen2020topology}
	Ruben Verresen.
	\href{https://arxiv.org/abs/2003.05453}{\newblock {\em  arXiv:2003.05453v1 [cond-mat.str-el]}} (2020).
	
	\bibitem{rahul2021majorana}
	S~Rahul, Ranjith~R Kumar, Y R~Kartik, and Sujit Sarkar.
	\href{https://journals.jps.jp/doi/abs/10.7566/JPSJ.90.094706?journalCode=jpsj}{\newblock {\em J. Phys. Soc. Jpn.} \textbf{90}, 094706 (2021).}
	
	\bibitem{niu2021emergent}
	Sen Niu, Yucheng Wang, and Xiong-Jun Liu.
	\href{https://arxiv.org/abs/2106.13400}{\newblock {\em arXiv:2106.13400v2 [cond-mat.str-el]}} (2021).
	
	
	\bibitem{PhysRevB.104.075132}
	Ryan Thorngren, Ashvin Vishwanath, and Ruben Verresen.
	\href{https://journals.aps.org/prb/abstract/10.1103/PhysRevB.104.075132}{\newblock {\em Phys. Rev. B.} \textbf{104}, 075132 (2021).}
	
	\bibitem{PhysRevResearch.3.043048}
	Oleksandr Balabanov, Daniel Erkensten, and Henrik Johannesson.
	\href{https://journals.aps.org/prresearch/abstract/10.1103/PhysRevResearch.3.043048}{\newblock {\em Phys. Rev. Res.} \textbf{3}, 043048 (2021).}
	
	\bibitem{fraxanet2021topological}
	Joana Fraxanet, Daniel Gonz{\'a}lez-Cuadra, Tilman Pfau, Maciej Lewenstein, Tim
	Langen, and Luca Barbiero.
	\href{https://journals.aps.org/prl/abstract/10.1103/PhysRevLett.128.043402}{\newblock {\em Phys. Rev. Lett.} \textbf{128}, 043402 (2022).}
	
	\bibitem{keselman2015gapless}
	Anna Keselman, Erez Berg.
	\href{https://journals.aps.org/prb/abstract/10.1103/PhysRevB.91.235309}{\newblock {\em Phys. Rev. B.} \textbf{91}, 235309 (2015).}
	
	\bibitem{scaffidi2017gapless}
	Thomas Scaffidi, Daniel E. Parker, Romain Vasseur.
	\href{https://journals.aps.org/prx/abstract/10.1103/PhysRevX.7.041048}{\newblock {\em Phys. Rev. X.} \textbf{7}, 041048 (2017).}
	
	\bibitem{duque2021topological}
	Carlos M. Duque, Hong-Ye Hu, Yi-Zhuang You, Vedika Khemani, Ruben Verresen, and Romain Vasseur.
	\href{https://journals.aps.org/prb/abstract/10.1103/PhysRevB.103.L100207}{\newblock {\em Phys. Rev. B.} \textbf{103}, L100207 (2021).}
	
	\bibitem{kumar2021multi}
	Ranjith~R Kumar, Y R~Kartik, S~Rahul, and Sujit Sarkar.
	\href{https://www.nature.com/articles/s41598-020-80337-7}{\newblock {\em Sci. Rep.} \textbf{11}, 1--20 (2021).}
	
	\bibitem{PhysRevLett.42.1698}
	W.~P. Su, J.~R. Schrieffer, and A.~J. Heeger.
	\href{https://journals.aps.org/prl/abstract/10.1103/PhysRevLett.42.1698}{\newblock {\em Phys. Rev. Lett.} \textbf{42}, 1698--1701 (1979).}
	
	\bibitem{kitaev2001unpaired}
	A~Yu Kitaev.
	\href{https://iopscience.iop.org/article/10.1070/1063-7869/44/10S/S29}{\newblock {\em Phys. Usp.} \textbf{44}, 131 (2001).}
	
	\bibitem{hsu2020topological}
	Hsiu-Chuan Hsu and Tsung-Wei Chen.
	\href{https://journals.aps.org/prb/abstract/10.1103/PhysRevB.102.205425}{\newblock {\em Phys. Rev. B.} \textbf{102}, 205425 (2020).}
	
	\bibitem{niu2012majorana}
	Yuezhen Niu, Suk~Bum Chung, Chen-Hsuan Hsu, Ipsita Mandal, S~Raghu, and Sudip
	Chakravarty.
	\href{https://journals.aps.org/prb/abstract/10.1103/PhysRevB.85.035110}{\newblock {\em Phys. Rev. B.} \textbf{85}, 035110 (2012).}
	
	

	\bibitem{jackiw1976solitons}
	Roman Jackiw and Cl{\'a}udio Rebbi.
	\href{https://journals.aps.org/prd/abstract/10.1103/PhysRevD.13.3398}{\newblock {\em Phys. Rev. D.} \textbf{13}, 3398 (1976).}
	
	\bibitem{lu2011non}
	Jie Lu, Wen-Yu Shan, Hai-Zhou Lu, and Shun-Qing Shen.
	\href{https://iopscience.iop.org/article/10.1088/1367-2630/13/10/103016/meta}{\newblock {\em New J. Phys.} \textbf{13}, 103016 (2011).}
	
	\bibitem{shun2018topological}
	Shun-Qing Shen. 
	\href{https://link.springer.com/book/10.1007/978-3-642-32858-9}{\newblock {\em Topological Insulators: Dirac Equation in Condensed Matter}.} \newblock Springer (2018).
	
	\bibitem{PhysRevLett.115.177204}
	G.~Zhang and Z.~Song.
	\href{https://journals.aps.org/prl/abstract/10.1103/PhysRevLett.115.177204}{\newblock {\em Phys. Rev. Lett.} \textbf{115}, 177204 (2015).}
	
	\bibitem{kumar2022topological}
	Ranjith~R Kumar, Y R~Kartik, and Sujit Sarkar.
	\href{https://arxiv.org/abs/2211.03320}{\newblock {\em  arXiv:2211.03320 [cond-mat.str-el]}} (2022).
	
	\bibitem{chen2016scaling}
	Wei Chen.
	\href{https://iopscience.iop.org/article/10.1088/0953-8984/28/5/055601}{\newblock {\em J. Condens. Matter Phys.} \textbf{28}, 055601 (2016).}
	
	\bibitem{chen2016scalinginvariant}
	Wei Chen, Manfred Sigrist, and Andreas~P Schnyder.
	\href{https://iopscience.iop.org/article/10.1088/0953-8984/28/36/365501/meta}{\newblock {\em J. Condens. Matter Phys.} \textbf{28}, 365501 (2016).}
	
	\bibitem{chen2018weakly}
	Wei Chen.
	\href{https://journals.aps.org/prb/abstract/10.1103/PhysRevB.97.115130}{\newblock {\em Phys. Rev. B.} \textbf{97}, 115130 (2018).}
	
	\bibitem{chen2019universality}
	Wei Chen and Andreas~P Schnyder.
	\href{https://iopscience.iop.org/article/10.1088/1367-2630/ab2a2d}{\newblock {\em New J. Phys.} \textbf{21}, 073003 (2019).}
	
	\bibitem{molignini2018universal}
	Paolo Molignini, Wei Chen, and Ramasubramanian Chitra.
	\href{https://journals.aps.org/prb/abstract/10.1103/PhysRevB.98.125129}{\newblock {\em Phys. Rev. B.} \textbf{98}, 125129 (2018).}
	
	\bibitem{molignini2020generating}
	Paolo Molignini, Wei Chen, and R~Chitra.
	\href{https://journals.aps.org/prb/abstract/10.1103/PhysRevB.101.165106}{\newblock {\em Phys. Rev. B.} \textbf{101}, 165106 (2020).}
	
	\bibitem{abdulla2020curvature}
	Faruk Abdulla, Priyanka Mohan, and Sumathi Rao.
	\href{https://journals.aps.org/prb/abstract/10.1103/PhysRevB.102.235129}{\newblock {\em Phys. Rev. B.} \textbf{102}, 235129 (2020).}
	
	\bibitem{malard2020scaling}
	M~Malard, H~Johannesson, and W~Chen.
	\href{https://journals.aps.org/prb/abstract/10.1103/PhysRevB.102.205420}{\newblock {\em Phys. Rev. B.} \textbf{102}, 205420 (2020).}
	
	\bibitem{molignini2020unifying}
	Paolo Molignini, R~Chitra, and Wei Chen.
	\href{https://iopscience.iop.org/article/10.1209/0295-5075/128/36001}{\newblock {\em Europhys. Lett.} \textbf{128}, 36001 (2020).}
	
	\bibitem{rufo2019multicritical}
	Rufo, S., Lopes, N., Continentino, M.A. and Griffith, M.A.R.
	\href{https://journals.aps.org/prb/abstract/10.1103/PhysRevB.100.195432}{\newblock {\em Phys. Rev. B.} \textbf{100}, 195432 (2019).}
	
	\bibitem{kartik2021topological}
	Y R~Kartik, Ranjith~R Kumar, S~Rahul, Nilanjan Roy, and Sujit Sarkar.
	\href{https://journals.aps.org/prb/abstract/10.1103/PhysRevB.104.075113}{\newblock {\em Phys. Rev. B.} \textbf{104}, 075113 (2021).}
	
	
	\bibitem{PhysRevB.95.075116}
	Wei Chen, Markus Legner, Andreas R\"uegg, and Manfred Sigrist.
	\href{https://journals.aps.org/prb/abstract/10.1103/PhysRevB.95.075116}{\newblock {\em Phys. Rev. B.} \textbf{95}, 075116 (2017).}
	
	\bibitem{PhysRevLett.110.165304}
	Dmitry A. Abanin, Takuya Kitagawa, Immanuel Bloch, and Eugene Demler.
	\href{https://journals.aps.org/prl/abstract/10.1103/PhysRevLett.110.165304}{\newblock {\em Phys. Rev. Lett.} \textbf{110}, 165304 (2013).}
	
	\bibitem{duca2015aharonov}
	L. Duca, T. Li, M. Reitter, I. Bloch, M. Schleier-Smith, and U. Schneider.
	\href{https://www.science.org/doi/10.1126/science.1259052}{\newblock {\em Science.} \textbf{347}, 288--292 (2015).}
	
	\bibitem{PhysRevLett.90.227902}
	G. Vidal, J. I. Latorre, E. Rico, and A. Kitaev.
	\href{https://link.aps.org/doi/10.1103/PhysRevLett.90.227902}{\newblock {\em Phys. Rev. Lett.} \textbf{90}, 227902 (2003).}
	
	\bibitem{PhysRevLett.102.255701}
	Frank Pollmann, Subroto Mukerjee, Ari Turner, Joel E. Moore.
	\href{https://link.aps.org/doi/10.1103/PhysRevLett.102.255701}{\newblock {\em Phys. Rev. Lett.} \textbf{102}, 255701 (2009).}
	
	\bibitem{PhysRevLett.121.076802}
	Daniel Yates, Yonah Lemonik, Aditi Mitra.
	\href{https://link.aps.org/doi/10.1103/PhysRevLett.121.076802}{\newblock {\em Phys. Rev. Lett.} \textbf{121}, 076802 (2018).}

	\bibitem{PhysRevB.101.035109}
	Ziyu Tao, Tongxing Yan, Weiyang Liu, Jingjing Niu, Yuxuan Zhou, Libo Zhang, Hao
	Jia, Weiqiang Chen, Song Liu, Yuanzhen Chen, and Dapeng Yu.
	\href{https://journals.aps.org/prb/abstract/10.1103/PhysRevB.101.035109}{\newblock {\em Phys. Rev. B.} \textbf{101}, 035109, (2020).}
	
	\bibitem{niu2021simulation}
	Jingjing Niu, Tongxing Yan, Yuxuan Zhou, Ziyu Tao, Xiaole Li, Weiyang Liu, Libo
	Zhang, Hao Jia, Song Liu, Zhongbo Yan, et~al.
	\href{https://www.sciencedirect.com/science/article/pii/S2095927321001766}{\newblock {\em Sci. Bull.} \textbf{66}, 1168--1175 (2021).}
	
	\bibitem{goldman2016topological}
	Nathan Goldman, Jan~C Budich, and Peter Zoller.
	\href{https://www.nature.com/articles/nphys3803}{\newblock {\em Nat. Phys.} \textbf{12}, 639--645 (2016).}
	
	\bibitem{meier2016observation}
	Eric~J Meier, Fangzhao~Alex An, and Bryce Gadway.
	\href{https://www.nature.com/articles/ncomms13986}{\newblock {\em Nat. Commun.} \textbf{7}, 1--6 (2016).}
	
	\bibitem{an2018engineering}
	Fangzhao~Alex An, Eric~J Meier, and Bryce Gadway.
	\href{https://journals.aps.org/prx/abstract/10.1103/PhysRevX.8.031045}{\newblock {\em Phys. Rev. X.} \textbf{8}, 031045 (2018).}
	
	\bibitem{xie2019topological}
	Dizhou Xie, Wei Gou, Teng Xiao, Bryce Gadway, and Bo~Yan.
	\href{https://www.nature.com/articles/s41534-019-0159-6}{\newblock {\em npj Quantum Inf.} \textbf{5}, 1--5 (2019).}
	
	\bibitem{meier2018observation}
	Eric J. Meier, Fangzhao Alex An, Alexandre Dauphin, Maria Maffei,
	Pietro Massignan, Taylor L. Hughes, and Bryce Gadway.
	\href{https://www.science.org/doi/10.1126/science.aat3406}{\newblock {\em Science} \textbf{362}, 929--933 (2018).}
	
	\bibitem{jiang2011majorana}
	Liang Jiang, Takuya Kitagawa, Jason Alicea, AR~Akhmerov, David Pekker, Gil
	Refael, J~Ignacio Cirac, Eugene Demler, Mikhail~D Lukin, and Peter Zoller.
	\href{https://journals.aps.org/prl/abstract/10.1103/PhysRevLett.106.220402}{\newblock {\em Phys. Rev. Lett.} \textbf{106}, 220402 (2011).}
	
	\bibitem{kraus2012preparing}
	Christina~V Kraus, Sebastian Diehl, Peter Zoller, and Mikhail~A Baranov.
	\href{https://iopscience.iop.org/article/10.1088/1367-2630/14/11/113036}{\newblock {\em New J. Phys.} \textbf{14}, 113036 (2012).}
	
	\bibitem{malard2020multicriticality}
	Mariana Malard, David Brandao, Paulo~Eduardo de~Brito, and Henrik Johannesson.
	\href{https://journals.aps.org/prresearch/abstract/10.1103/PhysRevResearch.2.033246}{\newblock {\em Phys. Rev. Res.} \textbf{2}, 033246 (2020).}			
\end{thebibliography}
\end{document}